\documentclass{sig-alternate-2013}
\usepackage{color}
\usepackage{latexsym}
\usepackage{algorithm, algorithmic}
\usepackage{txfonts}
\usepackage{pstricks}
\usepackage{pst-tree}
\usepackage{soul}
\usepackage{multirow}
\usepackage{fancyvrb}
\usepackage{url}
\usepackage{listings}

\usepackage{balance}  

\usepackage[font={bf,small}]{caption}
\usepackage[skip=0cm,list=true,labelfont=it]{subcaption}

\usepackage{tikz}
\usetikzlibrary{matrix,positioning}
\tikzstyle{rblock} = [rectangle, draw, text width=5.5em, text centered, rounded corners, minimum height=1.5em]

\DefineVerbatimEnvironment{code}{Verbatim}{fontsize=\small}
\newcommand{\eat}[1]{}

\newcommand{\sep}{\ |\ }

\newtheorem{theorem}{Theorem}[section]
\newtheorem{lemma}[theorem]{Lemma}

\newtheorem{assumption}[theorem]{Assumption}

\newtheorem{definition}[theorem]{Definition}

\newtheorem{example}[theorem]{Example}

\newcommand{\referencenumber}{1}
\newtheorem{@examplereference}{Example \referencenumber \ignorespaces }

\newcommand{\trans}[1]{\llbracket #1 \rrbracket}
\newcommand{\tuple}[1]{\langle #1 \rangle}
\newcommand{\Q}{{\cal Q}} 
\newcommand{\T}{{\cal T}} 
\newcommand{\D}{{\cal D}} 

\begin{document}

\title{The Homeostasis Protocol: Avoiding Transaction Coordination Through Program Analysis}

\numberofauthors{1}
\author{
\begin{tabular}{p{17cm}}\centering
Sudip Roy$^{1,2\thanks{Work done while at Cornell University.}}$, Lucja Kot$^2$, Gabriel Bender$^{2}$, Bailu Ding$^2$,  Hossein Hojjat$^2$,  Christoph Koch$^3$, Nate Foster$^2$, Johannes Gehrke$^{2,4\footnotemark[1]}$
\\[1ex]
\end{tabular}\\
\vspace{-1ex} 
\begin{tabular}{p{3cm}}\centering
  \affaddr{$^1$Google Research}\\
\end{tabular}
\begin{tabular}{p{3cm}}\centering
  \affaddr{$^2$Cornell University}\\
\end{tabular}
\begin{tabular}{p{3cm}}\centering
  \affaddr{$^3$EPFL}\\
\end{tabular}
\begin{tabular}{p{3cm}}\centering
  \affaddr{$^4$Microsoft Corp.}\\
\end{tabular}\\
\begin{tabular}{p{17cm}}\centering
  \email{ \{sudip, lucja, gbender, blding, hojjat\}@cs.cornell.edu, christoph.koch@epfl.ch,\\ \{jnfoster, johannes\}@cs.cornell.edu}\\
\end{tabular}
}
 \makeatletter
 \let\@copyrightspace\relax
 \makeatother
  
\maketitle

\begin{abstract}
Datastores today rely on distribution and replication to achieve improved performance and fault-tolerance. But
correctness of many applications depends on strong consistency properties---something that can impose substantial
overheads, since it requires coordinating the behavior of multiple nodes. This paper describes a new approach to
achieving strong consistency in distributed systems while minimizing communication between nodes. The key insight is to
allow the state of the system to be inconsistent during execution, as long as this inconsistency is bounded and does not
affect transaction correctness. In contrast to previous work, our approach uses program analysis to extract semantic
information about permissible levels of inconsistency and is fully automated. We then employ a novel \emph{homeostasis
protocol} to allow sites to operate independently, without communicating, as long as any inconsistency is governed by
appropriate \emph{treaties} between the nodes. We discuss mechanisms for optimizing  
treaties based on workload characteristics to minimize communication, as well as a prototype implementation and
experiments that demonstrate the benefits of our approach on common transactional benchmarks. 
\end{abstract}

\newif\ifconferenceversion
\conferenceversionfalse 


\section{Introduction}\label{sec:intro}

Modern datastores are huge systems that exploit distribution and replication to achieve high availability and enhanced fault tolerance at scale. In many applications, it is important for the datastore to maintain global consistency to guarantee correctness. For example, in systems with replicated data, the various replicas must be kept in sync to provide correct answers to queries. More generally, distributed systems may have global consistency requirements that span multiple nodes. However, maintaining consistency requires coordinating the nodes, and the resulting communication costs can increase transaction latency by an order of magnitude or more \cite{Yu:2000:DEC:1251229.1251250}.

Today, most systems deal with this tradeoff in one of two ways. A popular option is to bias toward high availability and low latency, and propagate updates asynchronously. Unfortunately this approach only provides eventual consistency guarantees, so applications must use additional mechanisms such as compensations or custom conflict resolution strategies~\cite{Terry:1995:MUC:224056.224070, Dynamo}, or they must restrict the programming model to eliminate the possibility of conflicts~\cite{DBLP:conf/cidr/AlvaroCHM11}. Another option is to insist on strong consistency, as in Spanner \cite{Corbett:2012:SGG:2387880.2387905}, Mesa \cite{42851} or PNUTS \cite{PNUTS}, and accept slower response times due to the use of heavyweight concurrency control protocols.

This paper argues for a different approach. Instead of accepting a trade-off between responsiveness and consistency, we demonstrate that by carefully analyzing applications, it is possible to achieve the best of both worlds: strong consistency and low latency in the common case. The key idea is to exploit the semantics of the transactions involved in the execution of an application in a way that is safe and completely transparent to programmers.

It is well known that strong consistency is not always required to execute transactions correctly \cite{Kraska:2009:CRC:1687627.1687657,  Yu:2000:DEC:1251229.1251250}, and this insight has been exploited in protocols that allow transactions to operate on slightly stale replicas as long as the staleness is ``not enough to affect correctness''~\cite{Barbara-Milla:1994:DPT:615199.615201, Yu:2000:DEC:1251229.1251250}. This paper takes this basic idea much further, and develops mechanisms for automatically extracting safety predicates from application source code. Our \emph{homeostasis protocol} uses these predicates to allow sites to operate without communicating, as long as any inconsistency is appropriately governed. Unlike prior work, our approach is fully automated and does not require programmers to provide any information about the semantics of transactions.

\begin{figure}[t]
 \centering
 \includegraphics[width=0.8\columnwidth]{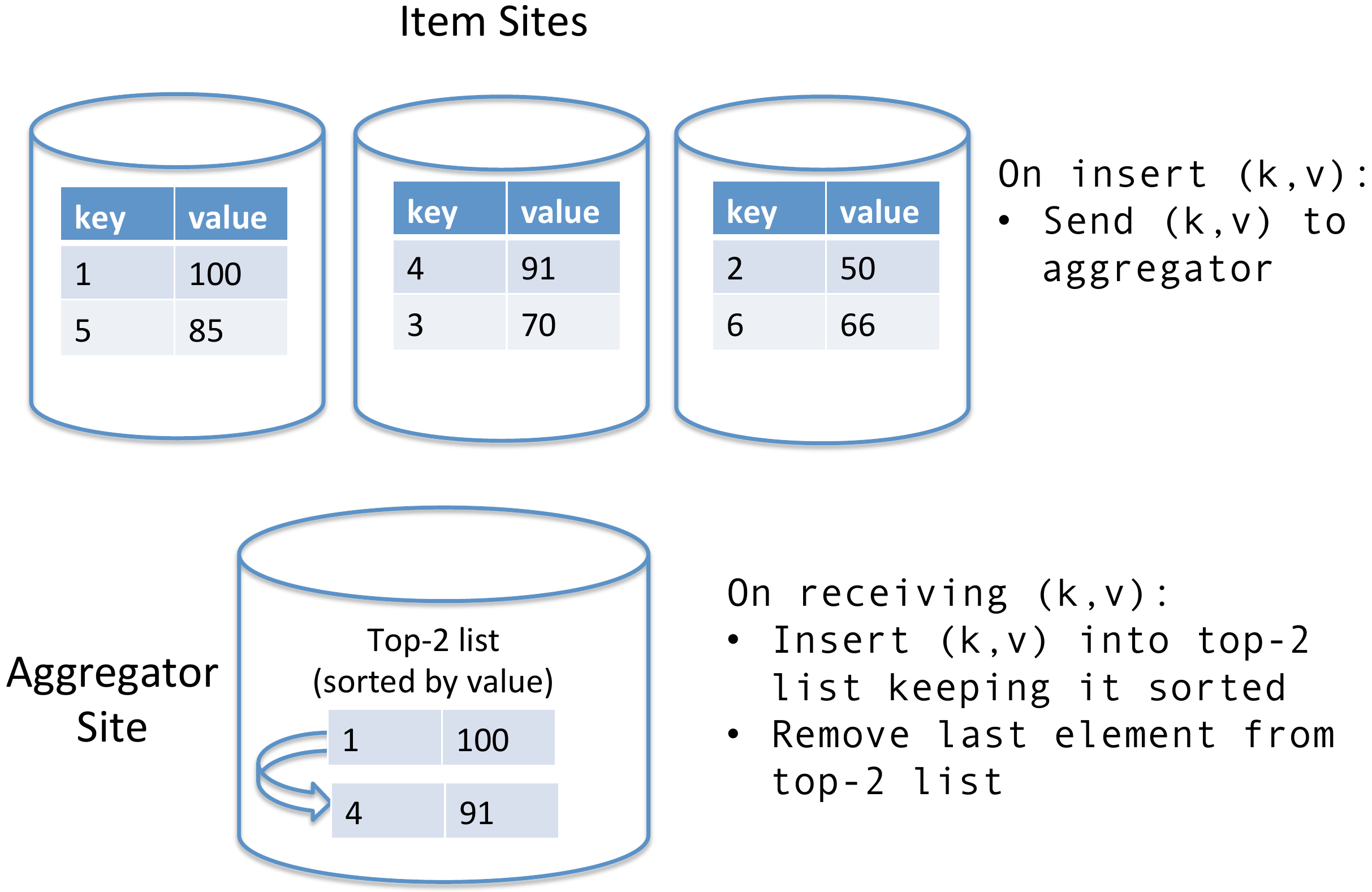}
 \caption{Distributed top-$2$ computation, basic algorithm. }
 \label{fig:topk_1}

\vspace{-3mm}

\end{figure}

\paragraph*{Example: top-$k$ query}
To illustrate the key ideas behind our approach in further detail, consider a top-$k$ query over a distributed datastore, as illustrated in Figure \ref{fig:topk_1}. For simplicity we will consider the case where $k = 2$. This system consists of a number of \emph{item sites} that each maintain a collection of $(\mathit{key}, \mathit{value})$ pairs that could represent data such as airline reservations or customer purchases. An \emph{aggregator site} maintains a list of top-$k$ items sorted in descending order by $value$. Each item site periodically receives new insertions, and the aggregator site updates the top-$k$ list as needed.

\begin{figure}[t]
 \centering
 \includegraphics[width=0.8\columnwidth]{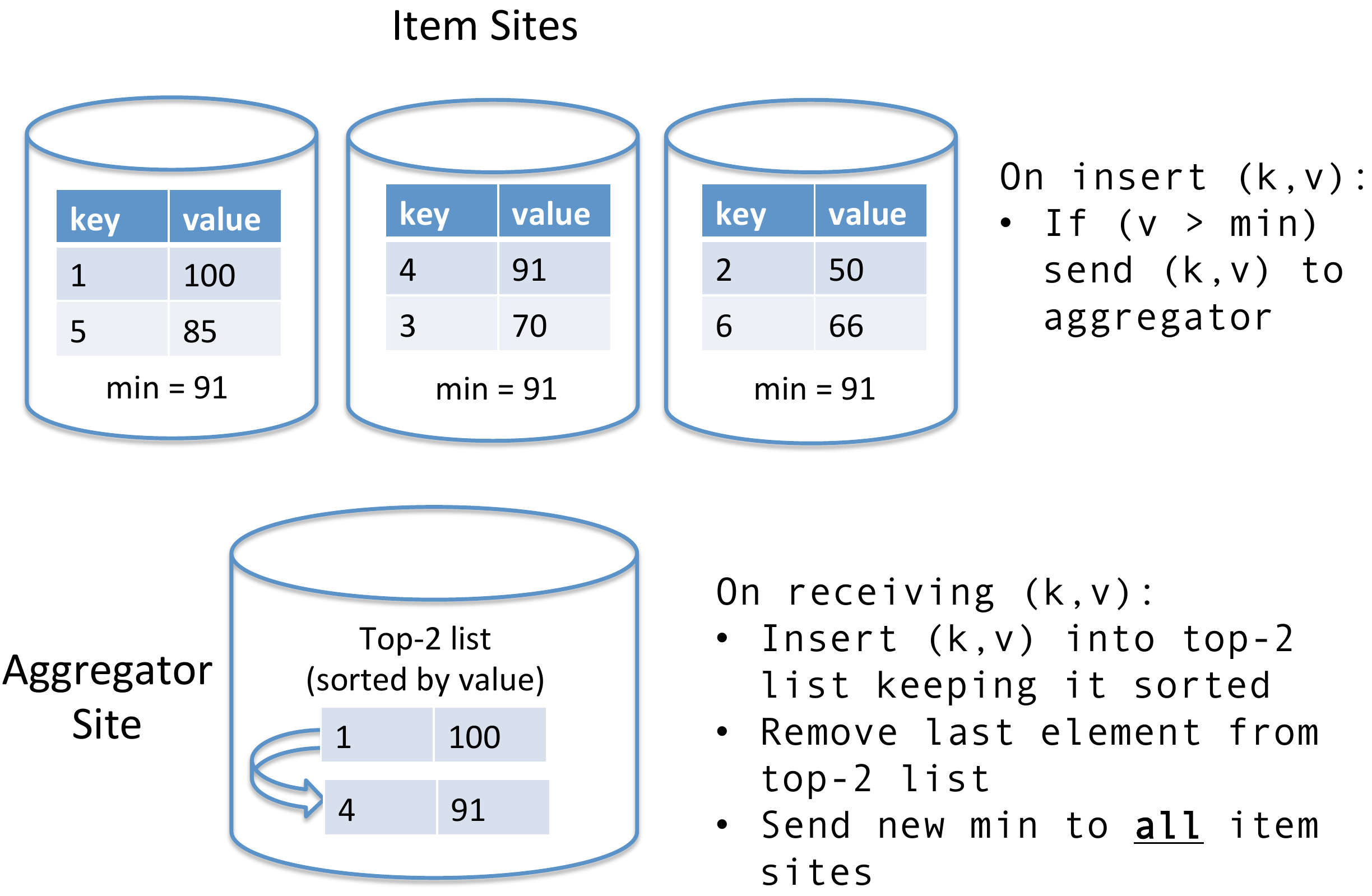}
 \caption{Distributed top-$2$ computation, improved algorithm. }
 \label{fig:topk_2}

\vspace{-4mm}

\end{figure}

A simple algorithm that implements the top-$k$ query is to have each item site communicate new insertions to the aggregator site, which inserts them into the current top-$k$ list in order, and removes the smallest element of the list. However, every insertion requires a communication round with the aggregator site, even if most of the inserts are for objects not in the top-$k$. A better idea is to only communicate with the aggregator node if the new $\mathit{value}$ is greater than the minimal value of the current top-$k$ list. Each site can maintain a cached value of the smallest value in the top-$k$ and only notify the aggregator site if an item with a larger value is inserted into its local state. This algorithm is illustrated in Figure \ref{fig:topk_2}, where each item site has a variable $\mathit{min}$ with the current lowest top-$k$ value. In expectation, most item inserts do not affect the aggregator's behavior, and consequently, it is safe for them to remain unobserved by the aggregator site.

This improved top-$k$ algorithm is essentially a simplified distributed version of the well-known \emph{threshold algorithm} for top-$k$ computation \cite{Fagin:2001:OAA:375551.375567}. However, note that this algorithm can be extracted \emph{automatically} by analyzing the code for the aggregator site.

\paragraph*{Our solution and contributions}
This paper presents a new semantics-based method for guaranteeing correct execution of transactions over an inconsistent datastore. Our method is geared towards improving performance in OLTP settings, where the semantics of transactions typically involve a shared global quantity (such as product stock level), most transactions make small incremental changes to this quantity, and transaction behavior is not sensitive to small variations in this quantity except in boundary cases (e.g. the stock level dips below zero). However, our method will still work correctly with workloads that do not satisfy these properties.

We make contributions at two levels. First, we introduce our solution as a high-level framework and protocol. Second, we give concrete instantiations and implementations for each component in the framework, thus providing one particular end-to-end, proof-of-concept system that implements our solution. We stress that the design choices we made in the implementation are not the only possible ones, and that the internals of each component within the overall framework can be refined and/or modified if desired.

%

The first step in our approach is to analyze the transaction code used in the application (we assume all such code is known up front) to compute a global predicate that tracks its ``sensitivity'' to different input databases. We then ``factorize'' the global predicate into a set of local predicates that can be enforced on each node. In the example above, the analysis would determine that the code for the aggregator site leaves the current top-$k$ list unchanged if each newly inserted item is smaller than every current top-$k$ element.

Analyzing transaction code is challenging. It is not obvious what the analysis should compute, although it should be some representation of the transaction semantics that explains how inputs affects outputs. Also, we need to design general analysis algorithms that are not restricted in the same ways as the demarcation protocol \cite{Barbara-Milla:1994:DPT:615199.615201} or conit-based consistency \cite{Yu:2000:DEC:1251229.1251250}, which are either limited to specific datatypes or require human input.

Our first contribution (Section \ref{sec:analysis}) is an analysis technique that describes transaction semantics using \emph{symbolic tables}. Given a transaction, a symbolic table is a set of pairs in which the first component is a logical predicate on database states and the second component concisely represents the transaction's execution on databases that satisfy the predicate. We show how to compute symbolic tables from transaction code in a simple yet expressive language.

Our second contribution (Section \ref{sec:protocol}) is to use symbolic tables to execute transactions while avoiding inter-node communication. Intuitively, we exploit the fact that if the database state does not diverge from the current entry in the symbolic table, then it is safe to make local updates without synchronizing, but if the state diverges then synchronization is needed. In our top-$k$ example in Figures \ref{fig:topk_1} and \ref{fig:topk_2}, as long as no site receives an item whose value exceeds the minimal top-$k$ value, no communication is required. On the other hand, if a value greater than the minimal value is inserted at a site then we would need to notify the aggregator, compute the new top-$k$ list, and communicate the new minimal value to all the sites in the system (including sites that have not received any inserts).

Our \emph{homeostasis protocol} generalizes the example just described and allows disconnected execution while guaranteeing correctness. To reason about transaction behavior in a setting where transactions are deliberately executed on inconsistent data, we introduce a notion of execution correctness based on \emph{observational equivalence} to a serial execution on consistent data. Our protocol is provably correct under the above definition, and subsumes the demarcation protocol and related approaches. It relies on the enforcement of \emph{global treaties}; these are logical predicates that must hold over the global state of the system. In our example, the global treaty is the statement ``the current minimal value in the top-$k$ is $91$''. If an update to the system state violates the treaty, communication is needed to establish a new treaty. 

In general, the global treaty can be a complex formula, and checking whether it has been violated may itself require inter-site communication. To avoid this, it is desirable to split the global treaty into a set of \emph{local treaties} that can be checked at each site and that together imply the global treaty. This splitting can be done in a number of ways, which presents an opportunity for optimization to tune the system to specific workloads. Our third contribution (Section \ref{sec:treaties}) is an algorithm for generating local treaties that are chosen based on known workload statistics. 

Implementing the homeostasis protocol is nontrivial and raises many systems challenges. Our fourth contribution is a prototype implementation and experimental evaluation (Sections \ref{sec:design} and \ref{sec:experiments}).

\section{Analyzing Transactions}
\label{sec:analysis}

This section introduces our program analysis techniques that capture the abstract semantics of a transaction in \emph{symbolic tables}. We first introduce our formal model of transactions (Sections \ref{subsec:transactions}) and symbolic tables (Section \ref{sec:symbolicqs}). We then present a simple yet expressive language $\mathcal{L}$ and explain how to compute symbolic tables for that language (Sections \ref{subsec:sem_tables}). Finally, we discuss a higher-level language $\mathcal{L}$++ (Section \ref{sec:lplusplus}) that builds on $\mathcal{L}$ without adding expressiveness and is more suitable for use in practice.
 
\subsection{Databases and transactions}\label{subsec:transactions}

We begin by assuming the existence of a countably infinite set \emph{Obj} of \emph{objects} that can appear in the database, denoted $x, y, z, \ldots$. A database can be thought of as a finite set of objects, each of which has an associated integer value. Objects not in the database are associated with a null default value. More formally, a database $D$ is a map from objects to integers that has finite support.

A \emph{transaction} $T$ is an ordered sequence of atomic \emph{operations}, which can be \texttt{read}s, \texttt{write}s, computations, or \texttt{print} statements. A \texttt{read} retrieves a value from the database, while a \texttt{write} evaluates an expression and stores the resulting value in a specific object in the database. A computation neither reads from nor writes to the database, but may update temporary variables. Finally, \texttt{print} statements allow transactions to generate user-visible output. We define transaction evaluation as follows:

\begin{definition}[Transaction evaluation]\label{def:xacteval}
The result of evaluating a transaction $T$ on a database $D$, denoted $\emph{Eval}(T, D)$, is a pair $\langle D', G' \rangle$ where $D'$ is an updated database and $G'$ is a log containing the list of values that the transaction printed during its execution. The order of values in the log is determined by the order in which the print statements were originally executed.
\end{definition}

We assume that transactions are \emph{deterministic}, meaning that $D'$ and $G'$ are uniquely determined by $T$ and $D$.

\subsection{Symbolic tables}\label{sec:symbolicqs}

We next introduce \emph{symbolic tables}, which allow us to capture the relationship between a transaction's input and its output. The symbolic table for a transaction $T$ can be thought of a map from databases to \emph{partially evaluated transactions}. For any database $D$, evaluating $T$ on $D$ is equivalent to first finding the transaction associated with $D$ in the symbolic table (which will generally be much simpler than $T$ itself; hence the term ``partially evaluated'') and then evaluating this transaction on $D$.

More formally, a symbolic table for a transaction $T$ is a binary relation $\Q_{T}$ containing pairs of the form $\langle \varphi_\D, \phi \rangle$ where $\varphi_\D$ is a formula in a suitably expressive first-order logic and $\phi$ is a transaction that produces the same log and final database state as $T$ on databases satisfying $\varphi_\D$. The symbolic tables for transactions $T_1$ and $T_2$ from Figure \ref{subfig:t1_L} and \ref{subfig:t2_L} are
shown in Figure \ref{subfig:q_t1} and Figure \ref{subfig:q_t2}. For example, the transaction $T_1$ behaves identically to the simpler transaction $\texttt{write}(x = \texttt{read}(x) + 1)$ on all databases which satisfy $x + y < 10$, so the symbolic table contains the tuple
\begin{equation*}
  \langle x + y < 10, \texttt{write}(x = \texttt{read}(x) + 1)\rangle
\end{equation*}

Typically, each execution path through the transaction code corresponds to a unique partially evaluated transaction, although this is not always true if, for example, a transaction executes identical code on both branches of a conditional statement. 

We can extend the notion of symbolic tables to work with a set of transactions $\{T_1, T_2, \cdots T_k\}$. A symbolic table for a set of $K$ transactions is a $K+1$-ary relation. Each tuple in this relation is now of the form $\langle \varphi_\D, \phi_1, ..., \phi_N \rangle$ where $\phi_i$ produces the same final database and log as $T_i$ when evaluated on any database satisfying $\varphi_\D$. Such a relation can be constructed from the symbolic tables of individual transactions as follows. For each tuple $\langle \varphi_1, \phi_1, ..., \varphi_N, \phi_N \rangle$ in the cross-product of the symbolic tables of all transactions we add a tuple $\langle \varphi_1 \wedge \varphi_2 \wedge ... \wedge \varphi_K, \phi_1, ..., \phi_N \rangle$ to the symbolic table for $\{ T_1, T_2, \ldots, T_k \}$. Figure \ref{subfig:q_t1t2} shows a symbolic table for  $\{T_1, T_2\}$.

\begin{figure}
 \begin{subfigure}[b]{0.24\textwidth}
 \begin{tabular}{cl}
  $T_1 ::= \{$& $\hat{x} := \texttt{read}(x);$\\
  & $\hat{y} := \texttt{read}(y);$\\
  & $\texttt{if}~(\hat{x} + \hat{y} < 10)~\texttt{then}$\\
  & ~~~ $\texttt{write}(x = \hat{x} + 1)$\\
  & $\texttt{else}$\\
  & ~~~ $\texttt{write}(x = \hat{x} - 1)$\\
  $~~~~\}$ &
 \end{tabular}
 \caption{Transaction $T_1$}
 \label{subfig:t1_L}
 \end{subfigure}
 ~
 \begin{subfigure}[b]{0.24\textwidth}
 \begin{tabular}{cl}
  $T_2 ::= \{$& $\hat{x}:=\texttt{read}(x);$\\
  & $\hat{y}:=\texttt{read}(y);$\\
  & $\texttt{if}~(\hat{x} + \hat{y} < 20)~\texttt{then}$\\
  & $~~~ \texttt{write}(y = \hat{y} + 1)$\\
  & $\texttt{else}$\\
  & $~~~ \texttt{write}(y = \hat{y} - 1)$\\
 $~~~~\}$  &
 \end{tabular}
 \caption{Transaction $T_2$}  
 \label{subfig:t2_L}
 \end{subfigure}
 
 \medskip
 
 \caption{Example transactions for symbolic tables. $x$ and $y$ are database objects, $\hat{x}$ and $\hat{y}$ are temporary program variables.}
 \label{fig:intensional_example}
\end{figure}

\begin{figure}

 \begin{subfigure}[b]{0.24\textwidth}
  \begin{center}
  \begin{tabular}{|c|c|} \hline
   $\varphi_\mathcal{D}$ & $\phi$ \\ \hline
   $x + y < 10$ & $\texttt{w}(x = \texttt{r}(x) + 1)$\\ \hline
   $x + y \geq 10$ & $\texttt{w}(x = \texttt{r}(x) - 1)$\\ \hline
  \end{tabular}
  \end{center}
  \vspace{-0.5em}
  
  \caption{Symbolic table for $T_1$.}
  \label{subfig:q_t1}
 \end{subfigure}
 ~
 \begin{subfigure}[b]{0.24\textwidth}
   \begin{center}
   \begin{tabular}{|c|c|} \hline
   $\varphi_\mathcal{D}$ & $\phi$ \\ \hline
   $x + y < 20$ & $\texttt{w}(y = \texttt{r}(y) + 1)$\\ \hline
   $x + y \geq 20$ & $\texttt{w}(y = \texttt{r}(y) - 1)$\\ \hline
   \end{tabular}
   \end{center}
   \vspace{-0.5em}
   
   \caption{Symbolic table for $T_2$.}
   \label{subfig:q_t2}
 \end{subfigure}

 \medskip
  
 \begin{subfigure}[b]{0.48\textwidth}
   \begin{center}
   \begin{tabular}{|c|c|c|} \hline
   $\varphi_\mathcal{D}$ & $\phi_1$ & $\phi_2$ \\ \hline
   $x + y < 10$ & $\texttt{w}(x = \texttt{r}(x) + 1)$ & $\texttt{w}(y = \texttt{r}(y) + 1)$\\ \hline
   $10 \leq x + y < 20 $ & $\texttt{w}(x = \texttt{r}(x) - 1]$ & $\texttt{w}(y = \texttt{r}(y) + 1)$\\ \hline
   $x + y \geq 20 $ & $\texttt{w}(x = \texttt{r}(x) - 1)$ & $\texttt{w}(y = \texttt{r}(y) - 1)$\\ \hline
   \end{tabular}
   \end{center}
   \vspace{-0.5em}
   \caption{Symbolic table for the transaction set $\{T_1, T_2\}$}
   \label{subfig:q_t1t2}
 \end{subfigure}
 
 \medskip
 
 \caption{Symbolic tables for $T_1$ and $T_2$. The \texttt{read} and \texttt{write} commands are abbreviated as \texttt{r} and \texttt{w} respectively.}
 \label{fig:quotient_multi}

\vspace{-4mm}
 
\end{figure}

\subsection{Computing symbolic tables}\label{subsec:sem_tables}

We now explain how to use program analysis to compute symbolic tables automatically from transaction code. Our analysis is specific to a custom language we call $\mathcal{L}$. This language is not Turing-complete -- in particular, it has no loops --  but it is  powerful enough to encode a wide range of realistic transactions. For instance, it allows us to encode all five TPC-C transactions \cite{tpcc}, which are representative of many realistic OLTP workloads. We describe how to express a large class of relational database queries in $\mathcal{L}$ in the Appendix, Section \ref{subsec:expressiveness}. 

Let $\widehat{Var}$ be a countably infinite set of temporary variables used in the transactions (and not stored in the database). Metavariables $\hat{x}, \hat{y}, \hat{i}, \hat{j}...$ range over temporary variables. There are four types of statements in $\mathcal{L}$:

\begin{itemize}
\addtolength{\itemsep}{-1.4ex}
\item arithmetic expressions \emph{AExp} (elements are denoted $e, e_0, e_1, ...$) 
\item boolean expressions \emph{BExp} (elements are denoted $b, b_0,   b_1, ...$)
\item commands \emph{Com} (elements are denoted $c, c_0, c_1, ...$)
\item transactions \emph{Trans} (elements are denoted $T, T_0, T_1, ..$). Each transaction takes a list \emph{Params} of zero or more integer parameters $p, p_0, p_1, ... $.
\end{itemize}

The syntax for the language is shown in Figure \ref{fig:L}. The expression $\texttt{read}(x)$, when evaluated, will return the current value of the database object $x$, while the command $\texttt{write}(x=e)$ will evaluate the expression $e$ and store the resulting value in database object $x$. There is also a \texttt{print} command that evaluates an expression $e$ and appends the resulting value to the end of a log. Unlike other commands, \texttt{print} produces output that is visible to external observers.

\begin{figure}
\begin{align*}
 (AExp)~~ e & ::=~ n \sep p \sep \hat{x} \sep e_0 \oplus e_1 \sep -e \sep \texttt{read}(x)\\
 (BExp)~~ b & ::=~ \texttt{true} \sep \texttt{false} \sep e_0 \odot e_1 \sep b_0 \wedge b_1 \sep \neg b \\
 (Com) ~~ c & ::=~ \texttt{skip} \sep \hat{x} := e \sep c_0; c_1 \sep \texttt{if}~b~\texttt{then}~c_1~\texttt{else}~c_2 \sep \\
 &~~~~~~~~\texttt{write}(x=e) \sep \texttt{print}(e)\\
 (Trans) ~~ T & ::=~ \{ c \} ~( P ) \\
 (Params) ~~ P & :=~ \texttt{nil} \sep p , P\\
 \oplus & ::= + \sep * ~~~~~~~~~
 \odot  ::= ~< \sep = \sep \leq 
\end{align*}
\caption{Syntax for the language $\mathcal{L}$}
\label{fig:L}
\end{figure}

Given a transaction $T$ in $\mathcal{L}$, we can construct a symbolic table for $T$   inductively using the rules shown in Figure \ref{fig:inductive}. $\varphi\{\dfrac{e}{x}\}$ denotes the formula obtained from $\varphi$ by substituting expression $e$ for all occurrences of $x$.

\begin{figure}[t]
 \centering

\begin{tikzpicture}[node distance = 1.0cm]
    \node [rblock] (readx) {$\hat{x}:=\texttt{read}(x)$};
    \matrix (table2) [right=.2cm of readx,matrix of nodes,row sep=0] {
		$x+y\ge10$   & $[\hat{x} \coloneqq \texttt{r}(x); \hat{y} \coloneqq \texttt{r}(y); \texttt{w}(x = \hat{x}-1)]$ \\\hline
		$x+y<10$       & $[\hat{x} \coloneqq \texttt{r}(x); \hat{y} \coloneqq \texttt{r}(y); \texttt{w}(x = \hat{x}+1)]$ \\
	};
    \node [rblock, below of=readx] (ready) {$\hat{y}:=\texttt{read}(y)$};
    \matrix (table3) [right=.2cm of ready,matrix of nodes,row sep=0] {
		$\hat{x}+y\ge10$   & $[\hat{y} \coloneqq \texttt{r}(y); \texttt{w}(x = \hat{x}-1)]$ \\\hline
		$\hat{x}+y<10$       & $[\hat{y} \coloneqq \texttt{r}(y); \texttt{w}(x = \hat{x}+1)]$ \\
	};
    \node [rblock, below of=ready] (ifcond) {$\texttt{if}(\hat{x}+\hat{y}<10)$};
    \matrix (table4) [right=.2cm of ifcond,matrix of nodes,row sep=0] {
		$\hat{x}+\hat{y}\ge10$   & $[\texttt{w}(x = \hat{x}-1)]$ \\\hline
		$\hat{x}+\hat{y}<10$       & $[\texttt{w}(x = \hat{x}+1)]$ \\
	};
    \node [rblock, below=.5cm of ifcond, node distance=1.5cm] (incx) {$\texttt{w}(x = \hat{x}+1)$};
    \node [rblock, right of=incx, xshift=1.5cm] (decx) {$\texttt{w}(x = \hat{x}-1)$};
    \matrix (table6) [below=0.01cm of incx,xshift=4.7cm,matrix of nodes,row sep=0] {
		$\textit{true}$   & $[\texttt{w}(x = \hat{x}-1)]$\\
	};
    \matrix (table5) [below=0.5cm of incx,xshift=4.7cm,matrix of nodes,row sep=0] {
		$\textit{true}$   & $[\texttt{w}(x = \hat{x}+1)]$\\
	};

    \coordinate (a) at (6,-5);
    \coordinate (b) at (6,0);
    \draw [->] (readx) -- (ready);
    \draw [->] (ready) -- (ifcond);
    \draw [->] (ifcond) -- (incx);
    \draw [->] (ifcond) -- (decx);
    \draw [->,dashed] (table2) -- (readx);
    \draw [->,dashed] (table3) -- (ready);
    \draw [->,dashed] (table4) -- (ifcond);
    \draw [->,dashed] (table5) -| (incx);
    \draw [->,dashed] (table6) -| (decx);
\end{tikzpicture}

 \caption{Example of symbolic table construction for $T_1$ from Figure \ref{subfig:t1_L}. Table construction occurs from bottom to top, and \texttt{read} and \texttt{write} are abbreviated as \texttt{r} and \texttt{w} respectively.}
 \label{fig:inductive_example}

\vspace{-4mm}

\end{figure}
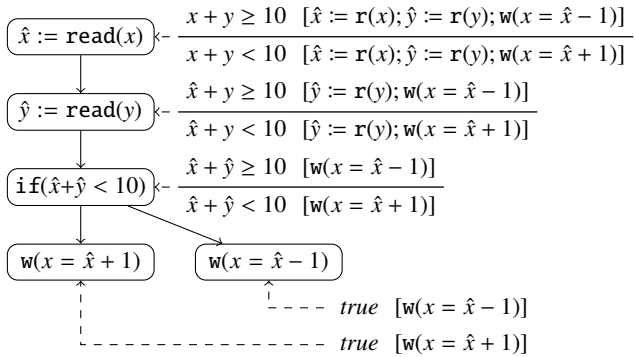

Algorithmically, we can compute the symbolic table by working backwards from the final statement in each possible execution path. Each such path generates one tuple in the symbolic table. We give an example of how to compute the symbolic table for transaction $T_1$ from Figure \ref{subfig:t1_L}. The left side of Figure \ref{fig:inductive_example} shows the code and the right side the computation.

\begin{figure}[t]
 \begin{align}
  \trans{T, \{\}} &\rightarrow \trans{c, \{\langle \texttt{true}, \texttt{skip}\rangle\}}~~~\text{ where $T = \{c\}$}\\
  \trans{c_1; c_2, \Q } &\rightarrow \trans{c_1, \trans{c_2, \Q}}\\
  \left\llbracket\begin{array}{l} \!\!\texttt{if}~b~\texttt{then}~c_1 \\ ~\texttt{else}~c_2  \end{array}\!\!, \Q \right\rrbracket &\rightarrow 
    \Bigg\{ \begin{array}{l}
    \left\{ \langle
    b \wedge \varphi, \phi \rangle~|~ \langle \varphi, \phi \rangle \in \trans{c_1, \Q} \right\} \cup\\
    \left\{ \langle
    \neg b \wedge \varphi, \phi \rangle~|~ \langle \varphi, \phi \rangle \in \trans{c_2, \Q} \right\}\end{array}\Bigg\}\\
  \trans{(\hat{x} \coloneqq e) ,\Q} &\rightarrow \Big\{ \langle
    \varphi\{\dfrac{e}{\hat{x}}\}, (\hat{x} \coloneqq e; \phi) \rangle~|~ \langle \varphi, \phi \rangle \in \Q
    \Big\}\\
  \trans{\texttt{skip}, \Q}  &\rightarrow \Q\\
  \trans{\texttt{write}(x=e), \Q} &\rightarrow \nonumber \\
    \Big\{ \langle &
    \varphi\{\dfrac{e}{x}\}, (\texttt{write}(x=e); \phi)\rangle ~|~ \langle \varphi, \phi
    \rangle \in \Q \Big\}\\
  \trans{\texttt{print}(e), \Q} &\rightarrow
    \Big\{ \langle
    \varphi, (\texttt{print}(e); \phi)\rangle ~|~ \langle \varphi, \phi
    \rangle \in \Q \Big\}
 \end{align}

 \caption{Rules for constructing a symbolic table for transaction $T$; $\Q$ is the running symbolic table.}
 \label{fig:inductive}
 \vspace{-4mm}
\end{figure}

First, Rule (1) initializes the symbolic table with a $true$ formula representing all database states and a trivial transaction consisting of a single \texttt{skip} command.
Next, the symbolic table is constructed by working backwards through the code as suggested by Rule (2). 
 
Because the last command is an \texttt{if} statement, we apply Rule (3) to create a copy of our running symbolic table; we use one copy for processing the ``true'' branch and the other for processing the ``false'' branch.  On each branch, we encounter a write to the variable $x$ and apply Rule (6). The statement immediately preceding the \texttt{if} is an assignment of a temporary variable. We apply Rule (4), performing a variable substitution on each formula and prepending a variable assignment to each partially evaluated transaction. We apply the same rule again to process the command $\hat{x} \coloneqq \texttt{read}(x)$.

Note that each tuple in the symbolic table corresponds to a unique execution path in the transaction, 
and thus a given database instance $D$ may only satisfy a single formula $\varphi$ in $\Q_T$. 

\subsection{From $\mathcal{L}$ to $\mathcal{L}$++}\label{sec:lplusplus}

$\mathcal{L}$ is a low-level language that serves to set up the foundation for our analysis; we do not expect that end users will specify transactional code in $\mathcal{L}$ directly. In particular, although it is possible to encode many SQL \texttt{SELECT} and \texttt{UPDATE} statements in $\mathcal{L}$ as described in the Appendix Section \ref{subsec:expressiveness}, the direct translation would lead to a large $\mathcal{L}$ program and an associated large symbolic table. Fortunately, we do not need to work with this $\mathcal{L}$ program directly. Because it was generated from a higher-level language, it will have a very regular structure which will translate into regularity in the symbolic table and enable compression.
 
Therefore, we have created a higher-level language we call $\mathcal{L}$++ which does not add any mathematical expressiveness to $\mathcal{L}$, but directly supports bounded arrays/relations with associated read, update, insert and delete operations. Our algorithm for computing symbolic tables for $\mathcal{L}$ programs can be extended in a straightforward way to compute symbolic tables for programs in $\mathcal{L}$++. 

\section{The Homeostasis Protocol}\label{sec:protocol}

This section gives the formal foundation for our homeostasis protocol that uses semantic information  to guarantee correct disconnected execution in a distributed and/or replicated system. We introduce a simplified system model (\ref{sec:system_model}) and several preliminary concepts (\ref{sec:lr}) and present the protocol (\ref{subsec:protocol_statement}). The material in this section applies to transactions in any language, not just $\mathcal{L}$ or $\mathcal{L}$++ from Section \ref{sec:analysis}.

\subsection{Distributed system model} \label{sec:system_model}

Assume the system has $K$ sites, each database object is stored on a single site, and transactions run across several sites. Formally, a distributed database is a pair $\langle D, \emph{Loc} \rangle$, where $D$ is as before and $\emph{Loc} : \emph{Obj} \rightarrow \{1, \hdots, K\}$ is a mapping from variables to the site identifiers where the variable values are located. Each transaction $T_j$ runs on a particular site $i$. Formally, there is a \emph{transaction location function} $\ell$ that maps transactions $T_j$ to site identifiers $\ell(T_j)$.

For ease of exposition, the discussion in Sections \ref{sec:lr} and \ref{subsec:protocol_statement} will focus on systems where the following assumption holds.

\begin{assumption}[All Writes Are Local]\label{assum:local}
If $T_j$ runs on site $i$, it only performs writes on variables $x$ such that $\emph{Loc}(x) = i$. In other words, all writes are local to the site on which the transaction runs.
\end{assumption}

This assumption can be lifted and we can apply the homeostasis protocol in systems and workloads with remote writes. We explain how and under what circumstances this can be done  in the Appendix, Section \ref{subsec:remotewrites}.

\subsection{LR-slices and treaties}\label{sec:lr}

As explained in the introduction, the idea behind the homeostasis protocol is to allow transactions to operate \emph{disconnected}, i.e. without inter-site communication, and over a database state that may be inconsistent, as long as appropriate \emph{treaties} are in place to bound the inconsistency and guarantee correct execution. Our protocol relies on the crucial assumption that all transaction code is known in advance. This assumption is standard for OLTP workloads.

In our model of disconnected execution, a transaction is guaranteed to read up-to-date values for variables that are local to the site on which it runs. For values resident at other sites, the transaction conceptually reads a locally available older -- and possibly stale -- \emph{snapshot} of these values. The actual implementation of how that snapshot is maintained can vary. If we are concerned about space overhead and do not wish to store an actual snapshot, we can simply transform the transaction code by inlining the snapshot value where it is needed. However, we will have to repeat this inlining at the start of each protocol round when the snapshots are synchronized with the actual remote values.

\begin{figure}
\hspace{-2mm}
 \begin{subfigure}[b]{0.24\textwidth}
 \begin{tabular}{cl}
  $T_3 ::= \{$& $\hat{x}:=\texttt{read}(x);$ \\
  &$ \texttt{if}~(\hat{x} > 0)~\texttt{then}$ \\
  & $~~~ \texttt{write}(y = 1)$\\
  & $\texttt{else}$ \\
  & $~~~ \texttt{write}(y = -1)$\\
  $~~~~\}$ &
 \end{tabular}
 \caption{Transaction $T_3$}
 \label{subfig:t5_L}
 \end{subfigure}
 \begin{subfigure}[b]{0.24\textwidth}
 \begin{tabular}{cl}
  $T_4 ::= \{$& $\hat{x}:=\texttt{read}(x); $\\
  &$\hat{y}:=\texttt{read}(y);$\\
  &$ \texttt{if}~(\hat{y} = 1)~\texttt{then} $\\
  &$ ~~~ \texttt{write}(z = (\hat{x} > 10))$\\
  &$ \texttt{else}$ \\
  &$ ~~~ \texttt{write}(z = (\hat{x} > 100))$\\
  $~~~~\}$ &
 \end{tabular}
 \caption{Transaction $T_4$}  
 \label{subfig:t6_L}
 \end{subfigure}


 \caption{Example transactions for protocol. $x$ is remote for both transactions, while $y$ and $z$ are local.}
 \label{fig:translations}

\vspace{-4mm}

\end{figure}

As a concrete example, consider transaction $T_3$ from Figure \ref{subfig:t5_L}. Suppose that $T_3$ runs on the site where $y$ resides, but that $x$ resides on another site. We can avoid the remote read of $x$ by setting $\hat{x}$ to a locally cached old value of $x$, say $10$. The transaction would still produce the correct writes on $y$ even if the value of $x$ should change on the remote site, as long as that site never allows $x$ to go negative. Thus, if we can set up a \emph{treaty} whereby the system commits to keeping $x$ positive, we can safely run instances of the transaction on stale data that is locally available.

To formalize the intuition provided by the above example, we explore how the local and remote reads of a transaction affect its behavior. Consider transaction $T_4$ from Figure \ref{subfig:t6_L}. Assume that $y$ and $z$ are local and $x$ is remote. The behavior of $T_4$ depends on both local and global state. Specifically, \emph{if the local value   $y$ is $1$} then it matters whether $x$ is greater than $10$, and \emph{if the local value $y$ is not $1$}, then it matters whether $x$ is greater than $100$.

The next few definitions capture this behavior precisely.

\begin{definition}[Local-remote partition]
Given a database $D$, a \emph{local-remote partition} is a boolean function $p$  on database objects in $D$. If $p(x) = true$ we say $x$ is \emph{local} under $p$; otherwise, it is \emph{remote}.
\end{definition}

Given a local-remote partition $p$, we can express any database $D$ as a pair $(l, r)$ where $l$ is a vector containing the values of local database objects and $r$ is a vector containing the values of remote objects. Any transaction $T$ has an associated local-remote partition that marks exactly the objects on site $\ell(T)$ as local.

We next formalize what it means for transactions to behave \emph{equivalently} on different databases. Our notion of equivalence is based on the \emph{observable execution} of transactions: we assume that an external observer can see the final database produced by a transaction as well as any logs it has produced (using \texttt{print} statements), but not the intermediate database state(s). Informally, the executions of two (sequences of) transactions are indistinguishable to an external observer if (i) both executions produce the same final database state and (ii) each transaction in the first sequence has a counterpart in the second sequence that produces exactly the same logs.

Alternative notions of observational equivalence are possible: for example, we could assume that an external observer can also see the intermediate database states as the transactions run. Although it is possible to define a variant of the homeostasis protocol that uses this stronger assumption, the protocol will allow fewer interleavings and less concurrency. We therefore use the weaker assumption that only the final database state and logs are externally observable; we believe this is well-justified in practice.

To formally define observational equivalence under  Assumption \ref{assum:local}, we do not need to worry about transactions' effect on the remote part of the database, because remote variables will not be written. This leads to the following definition. 

\begin{definition}[Observational equivalence ($\equiv$)]
Fix some local-remote partition. Let $l, l'$ be vectors of values to be assigned to local objects, let $r, r'$ be vectors of values to be assigned to remote objects, and let $G$ and $G'$ be logs. Then $\tuple{(l, r), G} \equiv \tuple{(l', r'), G'}$ if $l = l'$ and $G = G'$.
\end{definition}

The notion of equivalence defined above is reflexive, transitive, and symmetric. We can use it to formalize the assurances that our protocol provides when reading stale data.

\begin{definition}[Local-remote slice]\label{def:lrslice}
Let $T$ be a transaction with an associated local-remote partition. Consider a pair $L,R$, where each $l \in L$ is a vector of values to be assigned to local objects and each $r \in R$ is a vector of values to be assigned to remote objects. Then $\{(l, r)~|~ l \in L, r \in R \}$ defines a set of databases. Suppose $\emph{Eval}(T, (l, r)) \equiv \emph{Eval}(T, (l, r'))$ for every $l \in L$ and $r, r' \in R$. Then $(L, R)$ is a \emph{local-remote slice} (or LR-slice for short) for $T$.
\end{definition}

For any database in the LR-slice we know that as long as the remote values stay in $R$, the writes performed and the log produced by the transaction are determined \emph{only} by the local values.

\begin{example}
Consider our transaction $T_4$ from above, with $y$ being local and $x$ remote. For notational clarity, we omit $z$ and list the permitted values for $y$ first, then those for $x$. One LR-slice for $T_4$ is $( \{1\}, \{11, 12, 13\})$, another is $( \{1\}, \{11, 12, 13, 14\})$, and yet another is $( \{2, 3, 4\}, \{0, 1, 2, 3\})$.
\end{example}

There is no maximality requirement in our definition of $LR$-slices, as the above example shows. Next, we define global treaties.

\begin{definition}[Global Treaty]
A \emph{global treaty} $\Gamma$ is a subset of the possible database states $\mathcal{D}$ for the entire system. The system \emph{maintains} a treaty by requiring synchronization before allowing a transaction to commit with a database state that is not in $\Gamma$.
\end{definition}

For our purposes, it will be useful to have treaties that require the system to remain in the same $LR$-slice as the original database. Under such a treaty, by Definition \ref{def:lrslice}, it will be correct to run transactions that read old (and potentially out-of-date) values for remote variables. We formalize the treaty property we need as follows.

\begin{definition}[Valid Global Treaty]\label{def:valid_treaty}
Let $\{T_1, T_2, \ldots, T_n\}$ be a finite set of transactions. We say that $\Gamma$ is a $\emph{valid global treaty}$ if the following is true. For each transaction $T_j$ with associated site $\ell(T_j)$ and associated local-remote partition, let $L = \{l~|~(l, r) \in \Gamma\}$ and $R = \{r~|~(l, r) \in \Gamma\}$. Then $(L, R)$ must be a $LR$-slice for $T_j$.
\end{definition}

\subsection{Homeostasis protocol}\label{subsec:protocol_statement}

We next present our protocol to guarantee correct execution of transactions executed on inconsistent data. 

The protocol proceeds in rounds. Each round has three phases: treaty generation, normal execution and cleanup. Ideally, the system would spend most of its time in the normal execution phase of the protocol, where sites can execute transactions locally without the need for any synchronization. However, if a client issues a transaction $T'$ that would leave the database in a state that violates the current treaty, the system must end the current round of the protocol, negotiate a new treaty, and move on to the next round.

\medskip

{\bf Treaty generation:} In this phase, the system uses the current database $D$ to generate a valid global treaty $\Gamma$ such that $D \in \Gamma$.

\medskip

{\bf Normal execution:} In this phase, each site runs incoming transactions in a disconnected fashion, using local snapshots in reads of remote values as previously explained. Transactions running on a single site may interleave, as long as two invariants are enforced.

First, for each site $i$ there must exist a serial schedule $T_1, \ldots, T_m$ of the committed transactions at that site that produces the same log for each transaction and the same final database state as the interleaved schedule. This is a relaxation of  view-serializability and can be enforced conservatively by any classical algorithm that guarantees view-serializability. Second, executing any prefix $T_1, T_2, \ldots, T_j$ of the transactions in this sequence (with $j \leq m$) must yield a final database in $\Gamma$. This can be enforced by directly checking whether $\Gamma$ holds before committing each transaction. The details of how this check can be performed are discussed extensively in Section \ref{sec:treaties}.

If a pre-commit check ever determines that a running transaction $T'$ will terminate in a database state not in $\Gamma$, the transaction is aborted and the \emph{cleanup} phase begins.

\medskip

{\bf Cleanup:} First, the sites synchronize: each site $i$ broadcasts the value of every local object $x$ that has been updated since the start of the round. Next, we determine the transaction $T'$ that triggered the cleanup phase and was aborted, and we run $T'$ to completion at every site. Finally, we start a new round of the protocol.

\ifconferenceversion
\else 
It is crucial to ensure that synchronization occurs after all transactions accepted during the \emph{normal execution} phase have committed, and also that synchronization completes before $T'$ is executed on any site. $T'$ is deterministic and was known at the beginning of the cleanup phase. Consequently, executing $T'$ at every site after synchronization has completed ensures that $T'$'s writes are reflected on every site without the need for a second synchronization round.
\fi 

In a multi-site system, several transactions may try to make treaty-violating writes at the same time; we rely on a suitable voting algorithm to choose one ``winner'' $T'$ among them. We abort any ``loser'' transactions that also tried to violate the treaty and wait for all transactions other than $T'$ to terminate. (If, due to the local concurrency control algorithm used, some transactions are blocking/waiting for $T'$, they are aborted as well.). At the start of the next round, we rerun any remaining ``loser'' transactions.

\bigskip

We now argue that the protocol above is correct. It is not trivial to capture a good notion of ``correctness;'' we are after all reasoning about transactions that execute on \emph{inconsistent data}, which is not a common setting in classical concurrency theory.

Our notion of correctness is based on observational equivalence as introduced in Section \ref{sec:lr}. Even though transactions are permitted to operate on inconsistent data, it should be impossible for an external observer to distinguish between the execution of the transactions $T_1, T_2, \ldots, T_n$ using the homeostasis protocol and a serial execution of those same transactions that started on a consistent database. In particular, this means that (i) every transaction executed by the protocol must produce the ``right'' log, and (ii) we must end up with the ``right'' final database at the end of each round.

The following theorem captures the correctness of our protocol following the above intuition. In the theorem statement, we use the fact that the transactions executed at any given site can be totally ordered without altering the final database or logs that are produced -- this is guaranteed by the invariants that the protocol enforces during the \emph{normal execution} phase (discussed above).

\begin{theorem} \label{thm:protoocol-correctness}
Suppose that $\Gamma$ is a valid global treaty, and let $T_1, T_2, \ldots, T_n$ be a total order on the transactions executed during a single round of the homeostasis protocol such that if $T_j$ is executed before $T_k$ on some site $i$ then $j < k$. If we execute the transactions $T_1, T_2, \ldots, T_n$ in sequence then (i) each $T_j$ produces the same logs as it did in the homeostasis protocol, and (ii) the final database state is the same as the one produced by the homeostasis protocol.
\end{theorem}

\ifconferenceversion
The proof is in the technical report version of our paper \cite{homeostasis_tech_report}.
\else 
\begin{proof}
Notice that $T_n$ is the unique transaction that is executed for the first time at every site during the homeostasis protocol's \emph{cleanup} phase. Each of the remaining $n-1$ transactions is run for the first time during the \emph{normal execution} phase of the protocol.

We introduce variables $D_0, D_1, \ldots, D_n$ that represent the database state at different points in the serial execution of $T_1, T_2, \ldots, T_n$. We define $D_0$ to be the initial database at the start of the round, and for each $j = 1, 2, \ldots, n$ we define $D_j$ to be the database obtained by executing transactions $T_1, T_2, \ldots, T_j$ in sequence.
 
We also introduce variables that represent the state of the database at different sites during the protocol's execution. For each site $i$ and each $j = 0, 1, 2, \ldots, n - 1$ we define $D^i_k$ to be the database on site $i$ after executing every transaction $T_j$ such that $j \leq k$ and $\ell(T_j) = i$. Finally, for each site $i$ we define $D^i_n$ to be the database at site $i$ after the \emph{cleanup} phase has finished. It may be helpful to imagine a single client issuing the transactions $T_1, T_2, \ldots, T_{n-1}, T_n$ in sequence, sending each transaction $T_j$ to site $\ell(T_j)$ and waiting until each transaction has completed before issuing the next. Under this interpretation, $D^i_k$ is the database at site $i$ just after the $k$th transaction has finished.

When the transactions $T_1, T_2, \ldots, T_n$ are executed in sequence on the initial database $D_0$, we define $G_j$ to be the log produced by the $j$th transaction for each $j = 1, 2, \ldots, n$. We also define $G'_j$ to be the log obtained when $T_j$ is run on site $\ell(T_j)$ during the homeostasis protocol for each $j = 1, 2, \ldots, n$.

Finally, we use the notation $\Pi_i(D)$ to denote the restriction of a database $D$ to variables located at site $i$.

We must show that (i) each $T_j$ produces the same log in the serial execution as in the distributed protocol and (ii) the final database states are the same in both cases. Formally, we claim that $G_j = G'_j$ for each $j = 1, 2, \ldots, n$ and that $D_n = D^i_n$ for each site $i$.

\bigskip

We begin by considering transactions run during the \emph{normal execution} phase of the protocol. For each $j = 1, 2, \ldots, n-1$ and each site $i$, we claim that $\Pi_i(D^i_j) = \Pi_i(D_j)$ and that $G_j = G'_j$. The proof is by induction on $j$. We may assume that $\Pi_i(D^i_{j-1}) = \Pi_i(D_{j-1})$ for each site $i$. If $j = 1$, this is true because $D_0 = D^i_0$. If $j > 1$, it is true by the inductive hypothesis. Under this assumption, we must show that $\Pi_i(D^i_j) = \Pi_i(D_j)$ for every site $i$. Define $i' = \ell(T_j)$. If $i = i'$ then $G_i = G'_i$ and $\Pi_i(D^i_j) = \Pi_i(D_j)$ because $\Gamma$ is a valid global treaty. If $i \neq i'$ then $D^i_j = D^i_{j-1}$, so that $\Pi_i(D^i_j) = \Pi_i(D^i_{j-1})$. Furthermore, $\Pi_i(D_j) = \Pi_i(D_{j-1})$ because all writes performed by $T_j$ are local. It follows that $\Pi_i(D^i_j) = \Pi_i(D^i_{j-1}) = \Pi_i(D_{j-1}) = \Pi_i(D_j)$.

\bigskip

We next consider what happens in the transaction's \emph{cleanup} stage. The sites synchronize at the beginning of this stage. After synchronization is complete, each site has a database that agrees with $D^i_{n-1}$ on variables $x$ such that $\emph{Loc}(x) = i$. Since $\Pi_i(D^i_{n-1}) = \Pi_i(D_{n-1})$, this means the database at every site is exactly $D_{n-1}$. Finally, we run $T_n$ at every site $i$; this produces the log $G'_n = G_n$, and the resulting database is $D^i_n = D_n$.
\end{proof}

It is interesting to note that Assumption \ref{assum:local} does not necessarily need to hold of the unique transaction $T_n$ that is executed during the homeostasis protocol's \emph{cleanup} phase. In fact, $T_n$ can read and write \emph{arbitrary} database variables; for the proof above, we require only that it be deterministic. In the worst case, the protocol will execute only one transaction each round, but this transaction can perform \emph{arbitrary} reads and writes. The homeostasis protocol is reduced to a very expensive distributed sequencer in this case.
\fi 

\section{Generating treaties}\label{sec:treaties}

Having laid the theoretical groundwork for the homeostasis protocol, we move to the two most important practical considerations: how the global treaty is generated and how it is enforced. Our approach uses  symbolic tables as described in Section \ref{sec:analysis}; however, it does not depend on the way that symbolic tables are computed, or even on the language of the transactions. It can be applied to any workloads for which we can compute symbolic tables.

As a  running example, we use transactions $T_1$ and $T_2$ from Figure \ref{fig:intensional_example}. Assume that $x$ and $y$ are on different sites, $T_1$ runs on the site where $x$ resides, and $T_2$ runs on the site where $y$ resides.

For ease of exposition, we temporarily assume that we are working with transactions whose ``output'' (writes and logs) are functions of local database objects only; formally:

\begin{assumption}\label{assum:noremoteread}
For each transaction $T$, each partially evaluated transaction $\phi_T$ in the symbolic table for $T$, and each database variable $x$ that is read by $\phi_T$ we assume that $Loc(x) = \ell(T)$.
\end{assumption} 

This assumption holds for $T_1$ and $T_2$ from our previous example; we explain how to lift it in the Appendix, Section \ref{sec:localmisc}.

\subsection{Global and local treaties}\label{subsec:global_treaty}

In the homeostasis protocol, the first step in every round is to generate a valid global treaty (per Definition \ref{def:valid_treaty}). A na\"ive but valid global treaty would restrict the database state to the initial $D$, i.e. require renegotiation on every write. This reduces the homeostasis protocol to a distributed locking protocol; of course, we hope to do better. Intuitively, a good global treaty is one that maximizes the expected length of a protocol round---i.e. one that has a low probability of being violated.

To find a good global treaty, we use the symbolic tables we computed for our transaction workload. At the start of a protocol round, we know the current database $D$, the set $\T$ of all transactions that may run in the system, and the symbolic table $\Q_\T$ for $\T$, as shown in Figure \ref{subfig:q_t1t2} for our example  set $\{T_1, T_2\}$.

We then pick the unique formula $\psi$ in $\Q_\T$ that is satisfied by $D$. For $T_1$ and $T_2$, assume that the initial database state is $x = 10$ and $y = 13$. Then $\psi$ is $x + y \geq 20$ (third row of $\Q_\T$ in Figure \ref{subfig:q_t1t2}). Note that the symbolic table can be precomputed once and reused at each treaty computation, so finding $\psi$ can be done very quickly.

Unfortunately, $\psi$ is not guaranteed to be a valid global treaty under Definition \ref{def:valid_treaty}. In addition, even if it were, it would not be very useful to us as is, since it may be an arbitrarily complex formula and checking whether the database state satisfies $\psi$ may itself require inter-site communication.

For these two reasons, our next step is to factorize $\psi$ into a collection of \emph{local treaties} $\langle \varphi_{\Gamma_1}, \varphi_{\Gamma_2}, \ldots, \varphi_{\Gamma_K} \rangle$. Each local treaty $\varphi_{\Gamma_i}$ is a first-order logic formula associated with the site $i$ and uses as free variables only the database objects on site $i$. The conjunction of all the local treaties should imply $\psi$:
\begin{equation} \tag{H1} \label{horn1}
  \forall D'. \bigwedge_{1\le i\le K} \varphi_{\Gamma_i}(D') \rightarrow \psi(D')
\end{equation}
We also require each local treaty to hold on the original database $D$; this excludes trivial factorizations such as setting all local treaties to ${\it false}$. Formally, for each $i$ ($1\le i\le K$) we require the following:
\begin{equation} \tag{H2} \label{horn2}
  \varphi_{\Gamma_i}(D)
\end{equation}

\noindent Finally, we take the conjunction of all the local treaties to obtain a global treaty:
\begin{equation*}
  \varphi_\Gamma = \bigwedge_{1 \le i \le K} \varphi_{\Gamma_i}(D')
\end{equation*}

\noindent Due to the way the global treaty is constructed, all sites can enforce only their respective local treaties and still have the assurance that the global treaty is preserved.

\begin{lemma}\label{lemma:valid_treaty}
Let $\Gamma$ be the set of all databases satisfying $\varphi_{\Gamma}$. Under Assumption \ref{assum:noremoteread}, $\Gamma$ is a valid global treaty.
\end{lemma}

\ifconferenceversion
The proof is in the technical report version of our paper \cite{homeostasis_tech_report}.
\else 
\begin{proof}
Let $T$ be a transaction on site $i$ with local-remote partition $p_i$, and define $L = \{l~|~(l, r) \in \Gamma\}$ and $R = \{r~|~(l, r) \in \Gamma\}$. To show that $\Gamma$ is a valid global treaty, we must prove that $(L, R)$ is a $LR$-slice. More concretely, we must show that if $(l, r)$, $(l', r')$, and $(l'', r'')$ are all databases in $\Gamma$ then $\emph{Eval}(T, (l, r')) \equiv \emph{Eval}(T, (l, r''))$. It suffices to show that $\emph{Eval}(T, (l, r')) \equiv \emph{Eval}(T, (l, r)) \equiv \emph{Eval}(T, (l, r''))$ because equivalence is transitive. We will verify the first equivalence explicitly; since equivalence is symmetric, the second follows from an identical argument.

We first claim that $(l, r') \in \Gamma$. To prove this, we must check that $\varphi_{\Gamma_{i'}}(l, r')$ holds for every site $i' = 1, 2, \ldots, K$. If $i = i'$ then $\varphi_{\Gamma_{i'}}$ depends only on variables in $l$, so $\varphi_{\Gamma_{i'}}(l, r') = \varphi_{\Gamma_{i'}}(l, r)$, and $\varphi_{\Gamma_{i'}}(l, r)$ holds because $(l, r) \in \Gamma$. On the other hand, if $i \neq i'$ then $\varphi_{\Gamma_{i'}}$ depends only on variables in $r'$, so that $\varphi_{\Gamma_{i'}}(l, r') = \varphi_{\Gamma_{i'}}(l', r')$, and $\varphi_{\Gamma_{i'}}(l', r')$ holds because $(l', r') \in \Gamma$. Since $\varphi_{\Gamma_{i'}}(l, r')$ holds for every $i'$, the hypothesis follows.

We now know that $(l, r) \in \Gamma$ (by assumption) and that $(l, r') \in \Gamma$ (by the argument above). Hence, $(l, r)$ and $(l, r')$ are both associated with the same element $\tuple{\varphi, \phi_T}$ of the symbolic table for $T$, so that
\begin{align*}
              \emph{Eval}(T, (l, r)) & = \emph{Eval}(\phi_T, (l, r)) \\
  \text{and } \emph{Eval}(T, (l, r')) & = \emph{Eval}(\phi_T, (l, r'))
\end{align*}
both hold. Under Assumption \ref{assum:noremoteread}, $\phi_T$ will perform the same writes when executed on the database $(l, r)$ as when executed on $(l, r')$, so that $\emph{Eval}(\phi_T, (l, r)) \equiv \emph{Eval}(\phi_T, (l, r'))$. Combining these two observations, we conclude that $\emph{Eval}(T, (l, r)) \equiv \emph{Eval}(T, (l, r'))$, as desired.
\end{proof}
\fi 

From now on, we will blur the distinction between $\Gamma$ and $\varphi_{\Gamma}$, and work with an intensional notion of a treaty as a  first-order logic formula that represents a set of possible database states.

\subsection{Computing good local treaties}\label{sec:optimization}

The crux to computing $\varphi_{\Gamma}$ is finding local treaties, which is difficult in the general case. The requirements  in~\ref{horn1} and~\ref{horn2} create a set of Horn clause constraints $\mathcal{H}$. For an undecidable fragment of first-order logic (such as non-linear arithmetic over integers) finding the solution to $\mathcal{H}$ is undecidable, so the local treaties may not exist. For some first-order theories such as the theory of arrays, solving $\mathcal{H}$ is decidable but the local treaties are not necessarily quantifier-free. However, for certain theories such as linear integer arithmetic, the existence of solutions to $\mathcal{H}$ is decidable, solutions can be effectively computed and the solutions are guaranteed to be quantifier-free. These fragments of first-order logic are the theories that have the quantifier-free Craig's interpolation property. For a more elaborate discussion of solving Horn clause systems, see~\cite{RummerHK13}. 

We present a method for computing local treaties that is not precise -- i.e. it may not find the ``optimal'' local treaties that maximize protocol round length. However, it always works; in the worst case it degenerates to requiring synchronization on every write. It also allows us to optimize the local treaties based on a model of the expected future transaction workload. Such a model could be generated dynamically by gathering workload data as the system runs, or in other ways. The length of time spent in the optimization is a tunable knob so we can trade off treaty quality versus treaty computation time. This is important because local treaties need to be computed at the start of every protocol round; we do not want the cost of treaty computation to wipe out the savings from reduced communication during the protocol round itself.

There are three steps in our algorithm; preprocessing the global treaty, generating \emph{templates} for the local treaties, and running an optimization to instantiate the templates into actual local treaties.

We begin by preprocessing the formula $\psi$ to obtain a stronger but simpler formula that is a conjunction of \emph{linear constraints}. A linear constraint is an expression of the form
$$\left( \sum_i d_i x_i \right) \odot n$$
where the $d_i$ and $n$ are integers, the $x_i$ are variables and $\odot$ is either $=$, $<$ or $\leq$. The preprocessing causes a loss of precision in that we are now enforcing a stronger property than $\psi$. However, linear constraints and their variants arise frequently as global treaties in OLTP workloads and feature prominently in related work  \cite{Barbara-Milla:1994:DPT:615199.615201, Kraska:2009:CRC:1687627.1687657,  Yu:2000:DEC:1251229.1251250}; thus, we expect to handle most real-world global treaties with minimal precision loss. The preprocessing itself is straightforward and described in the Appendix, Section \ref{sec:localmisc}.

The next step is to  create \emph{local treaty templates}; 
these contain fresh \emph{configuration variables} that are instantiated later. We generate a template for each site $k$ using the following method. Start with $\psi$ and proceed clause by clause. Given a clause of the form

$$\left( \sum_i d_i x_i \right) \odot n,$$
we replace it by 
$$\left( \sum_{Loc(x_i) = k} d_i x_i + c_k \right) \odot n$$
where $c_k$ is a fresh configuration variable.

With our running example transactions $T_1$ and $T_2$, initial database state  $x = 10$ and $y = 13$ and $\psi: x + y \geq 20$, this process would yield the local treaty templates $\varphi_{\Gamma_1} :  (x + c_y \geq 20)$ and $\varphi_{\Gamma_2} :  (c_x + y \geq 20)$, where $c_x$ and $c_y$ are configuration variables. 

The following theorem states there is always at least one assignment of values to configuration variables that yields actual local treaties. We call any such assignment a \emph{valid treaty configuration}.

\begin{theorem}\label{thm:valid_config_exists}
Given a starting formula $\psi$ and database $D$, and a set of local treaty templates generated using the above process, there always exists at least one way to assign values to the configuration variables that satisfies~\ref{horn1} and \ref{horn2}.
\end{theorem}

\ifconferenceversion
The proof is in the technical report version of our paper \cite{homeostasis_tech_report}.
\else 
\begin{proof}
We generate the assignment of values to configuration variables as follows. Consider a configuration variable $c_k$ introduced to replace a reference to some variable $x$. If $c$ appears in an equality, i.e. a clause of the form
$$\sum_{Loc(x_i) = k} d_i x_i + c_k = n$$
then we replace it by $\sum_{Loc(x_j) \neq k} d_j D(x_j)$. This is an integer which can be computed by evaluating the expression above on the original database $D$.

Otherwise, if $c_k$ appears in an inequality of the form
$$\sum_{Loc(x_i) = k} d_i x_i + c_k \leq n$$
then we replace it with $- \sum_{Loc(x_i) = k} d_i D(x_i) + n$. Thus the expression simplifies to
$$\sum_{Loc(x_i) = k} d_i x_i - \sum_{Loc(x_i) = k} d_i D(x_i) + n \leq n$$
that is,
$$\sum_{Loc(x_i) = k} d_i x_i \leq  \sum_{Loc(x_i) = k} d_i D(x_i) $$

Finally, if $c_k$ appears in an inequality of the form
$$\sum_{Loc(x_i) = k} d_i x_i + c_k < n$$
then we replace it with $- \sum_{Loc(x_i) = k} d_i D(x_i) + n - 1$, which gives
$$\sum_{Loc(x_i) = k} d_i x_i - \sum_{Loc(x_i) = k} d_i D(x_i) + n - 1 < n$$
or
$$\sum_{Loc(x_i) = k} d_i x_i - 1 < \sum_{Loc(x_i) = k} d_i D(x_i) $$
or equivalently 
$$\sum_{Loc(x_i) = k} d_i x_i \leq \sum_{Loc(x_i) = k} d_i D(x_i) $$
which is the same expression as in the $\leq$ case.

\medskip

We now show the above is a valid treaty configuration. 

The first requirement is that the local treaties need to hold on $D$. This is straightforward because we know $\varphi_{\Gamma}$ holds on $D$. Observing the formulas we gave above (with the appropriate expressions for $c_k$ substituted in) makes it clear that all clauses of all local treaties must hold on $D$.

We now show that for any database $D'$, satisfying all the local treaties implies satisfying the global treaties. Suppose for a contradiction that all local treaties are satisfied on $D'$ but $\varphi_{\Gamma}$ does not hold on $D'$. Thus there must be at least one conjunct in $\varphi_{\Gamma}$ that is false on $D'$, although the corresponding conjunct is true in all the $\varphi_{\Gamma_i}$ on $D'$. We argue by cases, depending on whether the conjunct is an equality or inequality.

Start with the equality case. We know that the equality holds on the original $D$, so 
$$\sum_i d_i D(x_i) = n$$
Under the treaty configuration we have constructed, the local clause for site $k$ will have the following form:
$$\sum_{Loc(x_i) = k} d_i x_i + \sum_{Loc(x_j) \neq k} d_j D(x_j) = n$$
By assumption, the local clause at each site is true on the database $D'$, so that for each site $k$ the equality
$$\sum_{Loc(x_i) = k} d_i D'(x_i) + \sum_{Loc(x_j) \neq k} d_j D(x_j) = n$$
must hold. Summing over all sites yields the equality
$$\sum_i d_i D'(x_i) + (k-1) \left[\sum_i d_i D(x_i)\right] = kn$$
This is true because each variable is local to one site and remote to the $k-1$ other ones.
We can now use the fact that the expression is true on $D$, so that $\sum_i{d_i D(x_i)} = n$. Subtracting $(k-1) \left[\sum_i{d_i D(x_i)}\right]$ from both sides yields
$$\sum_i d_i D'(x_i) = n$$
giving the desired contradiction.

In the inequality case (both $<$ and $\leq$ as explained above), the local constraint that holds at site $k$ is 
$$\sum_{Loc(x_i) = k} d_i D'(x_i) \leq \sum_{Loc(x_i) = k} d_i D(x_i) $$
Summing these over all $k$ sites yields the following, as each variable appears exactly on one site:
$$\sum_i d_i D'(x_i) \leq \sum_i d_i D(x_i) $$
But by assumption the treaty holds on $D$ so  we know 
$$\sum_i d_i D(x_i) \leq n $$
and by transitivity 
$$\sum_i d_i D'(x_i) \leq n $$
giving the desired contradiction.
\end{proof}
\fi 

In general, many valid treaty configurations may exist. Our final step is to run an optimization to find a configuration that yields local treaties with a low probability of being violated. As mentioned, we assume we have a model that describes expected future workloads; the details of this model and how it is generated are independent of our algorithm. All that we need is a way to use the model to sample (generate) a number of possible future system executions. Once these are available, we create a MaxSAT instance and use a solver to generate the optimal treaty configuration for these future executions. The number and length of future executions we consider are tunable parameters that allow us to trade off quality of treaties versus time spent in the solver. The details of how we generate the MaxSAT instance are in the Appendix, Section \ref{sec:localmisc}. 

\subsection{Discussion and scope of our solution}\label{subsec:scope}

We have presented our approach for supporting disconnected yet correct transaction execution. The homeostasis protocol allows for provably correct disconnected execution as long as a \emph{global treaty} is maintained. The global treaty is enforced using \emph{local treaties} at each site. We compute the treaties automatically, by analyzing transaction code semantics to generate symbolic tables; treaty computation also takes into account expected workload characteristics.

Our key contributions are, first, a fully general homeostasis protocol, and second, a concrete instantiation and implementation of the protocol. Our instantiation uses the language $\mathcal{L}$++ for specifying transactions and performing analysis, and it uses the algorithm described in Section \ref{sec:optimization} for computing local treaties. However, it is possible to use the homeostasis protocol with another language and/or local treaty computation algorithm.

We now discuss the scope of the homeostasis protocol versus that of the most similar existing approaches such as the demarcation protocol \cite{Barbara-Milla:1994:DPT:615199.615201} and conit-based consistency \cite{Yu:2000:DEC:1251229.1251250}. Our current algorithm for local treaty computation only generates treaties which are conjunctions of linear constraints; the protocols in the above related work are able to enforce any such constraints if they are specified manually. However, generating linear constraints manually is tedious and error prone, and an automatic approach is preferable. 

Consider the problem of maintaining a more complex aggregate than top-$k$, such as a top-$k$ of minimums. This might arise in a weather monitoring scenario, where an application records temperature observations and displays the $k$ record daily low temperatures for a date, the $k$ \emph{highest} daily low temperatures, and so on. The constraints for correct disconnected computation of the top-$k$ daily low temperatures are linear and in principle derivable manually, but nontrivial. Suppose we now make the program more complex and also request the top-$k$ days with the highest temperature difference (between the daily low and high temperatures). The treaties required remain linear but become quite difficult to infer manually. We discuss both these examples in the Appendix, Section \ref{sec:linear_and_nonlinear}.

Moreover, there may be applications where the treaties cannot be expressed as linear constraints unless they are trivial (i.e., force synchronization on every write). In such cases, our algorithm from Section \ref{sec:optimization} will generate trivial treaties; however, our protocol is compatible with more powerful local treaty generation algorithms which are not so restricted, whereas the demarcation protocol is fundamentally limited to handle only linear constraints.

In summary, our automated approach removes the error-prone task of manually inferring constraints. Also, our protocol is fully independent of the language used for the transactions and for the treaties. Thus, our solution is open to handling more expressive treaties with suitable algorithmic extensions, which are future work.

\section{Homeostasis In Practice}\label{sec:design}
In this section, we highlight some of the practical challenges in
implementing the homeostasis protocol and discuss specific design
decisions we made in our prototype implementation.

\subsection{System design}\label{subsec:system}
Our system has two main components -- an offline component which analyzes application transaction code, and an online server which receives transaction execution requests from clients and coordinates with other servers to establish and maintain treaties. We discuss each of these components next.

\begin{figure}[t]
 \centering
 \includegraphics[width=0.9\columnwidth]{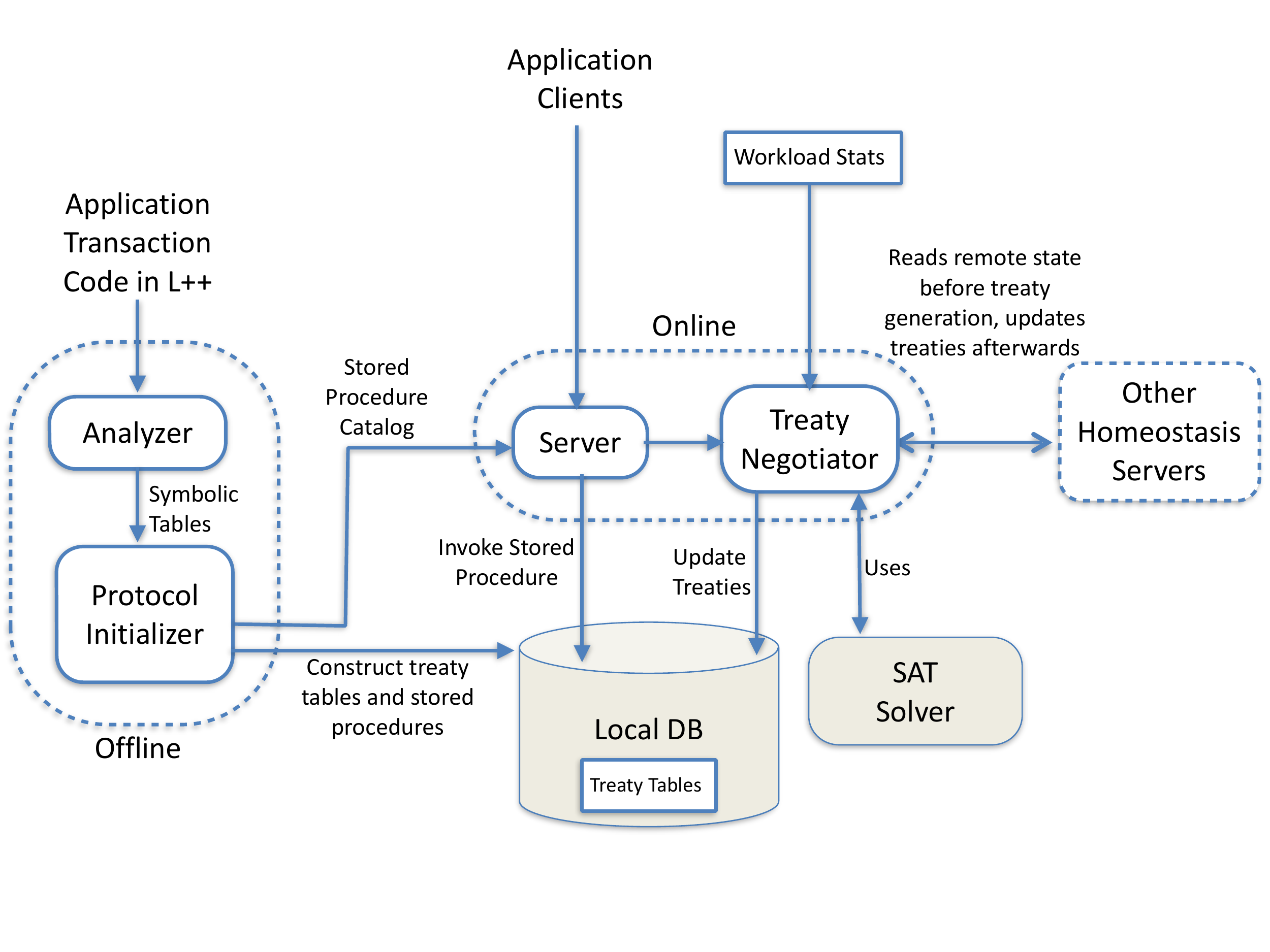}
 \caption{Homeostasis System Architecture.}
 \label{fig:system_architecture}
\vspace{-5mm}
\end{figure}

\textbf{Offline preprocessing} As shown in Figure \ref{fig:system_architecture}, there are two main offline components: the analyzer and the protocol initializer. 

The analyzer accepts as input transactions in $\mathcal{L}$++ (Section \ref{sec:lplusplus}) and computes (joint) symbolic tables. In doing so, it applies a number of compression techniques to exploit independence properties and keep the size of the symbolic tables small. 

Often transaction code operates on multiple database objects independently; for example,  the TPC-C New Order transaction orders several different items. The stock level of each item affects the transaction behavior, but each item affects a different portion of the code. Using a read-write dependency analysis like the one in SDD-1 \cite{Rothnie:1980:ISD:320128.320129}, we identify such points of independence and use them to encode symbolic tables more concisely in a factorized manner.

Moreover, transactions may take integer parameters, and the behavior of the transaction obviously depends on the concrete parameter values. Rather than instantiate parameters now, we push the parameterization into the symbolic tables for further compression.

The protocol initializer sets up the \emph{treaty table} -- a data structure that at any given time contains the current global treaty and the current local treaty configuration. The treaty table is thus dependent on the current database state; it is initialized offline based on the database state before the system starts accepting transaction requests. Subsequently, it is updated at each treaty negotiation in the online component.

The protocol initializer also performs some further setup for the online component. For every partially evaluated transaction in the symbolic tables produced by the analyzer, it creates and registers a stored procedure which executes this partially evaluated transaction. The stored procedure also includes checks for the satisfaction of the corresponding treaty as maintained in the treaty table. The stored procedure returns a boolean flag indicating whether the local treaty is violated after execution. The protocol initializer also creates a catalog that maps transactions to corresponding stored procedures in the treaty table.
 
\textbf{Online execution} The online component accepts and executes transactions using the homeostasis protocol.  When a transaction execution request arrives from the clients,  the system identifies the appropriate stored procedure in the catalog created during offline preprocessing. The server executes the stored procedure within the scope of a transaction. If the local treaty associated with the stored procedure is satisfied, then the transaction commits locally. Otherwise, the server invokes the treaty negotiator to synchronize with other servers and renegotiate a set of treaties. The negotiator uses an optimizer such as a SAT solver to determine local treaties. It then updates the treaty table and propagates the new treaties to all the other nodes. Therefore, every treaty negotiation requires two rounds of global 
communication---one for synchronizing database state across nodes and one for communicating the new treaties.  However, it is possible to eliminate the second round of communication if the solver is deterministic and therefore arrives at the same configuration at each of the replicas independently. 

Our implementation uses an underlying 2PC-like protocol for negotiation, and relies on the concurrency control mechanism of the transaction processing engine to ensure serializable execution locally. However, it would be easy to port it to any infrastructure which supports strongly consistent transaction execution.

\subsection{Implementation details}

Our system is implemented in Java as middleware built on top of the MySQL InnoDB Engine. Each system instance has a similar setup and communicates with the other instances through network channels. When handling failures, we currently rely on the recovery mechanisms of the underlying database. All in-memory state can be recomputed after failure recovery. In the offline component, we use ANTLR-4 to generate a parser for transactions in $\mathcal{L}$++. For finding optimal treaty configurations, we use the Fu-Malik Max SAT procedure \cite{Fu:2006:SPM:2165381.2165413} in the Microsoft Z3 SMT solver \cite{DeMoura:2008:ZES:1792734.1792766}.

\section{Evaluation}\label{sec:experiments}
We now show an experimental evaluation of a prototype implementation of the homeostasis protocol. 
We run a number of microbenchmarks (Section \ref{sec:micro}),
 as well as a set of experiments based on TPC-C \cite{tpcc}(Section \ref{sec:tpcc}) 
to evaluate our implementation in a realistic setting. 
All experiments run in a replicated system, using the transformations described in the Appendix, Section \ref{subsec:remotewrites}.

\begin{figure*}[th]
\begin{minipage}[t]{0.3\textwidth}
    \centering
    \includegraphics[width=\textwidth]{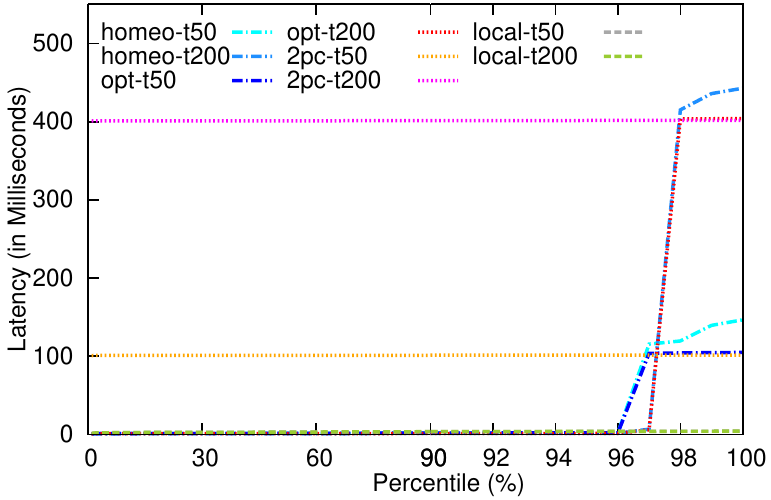}
    \caption{Latency with network RTT ($N_r=2, N_c=16$)}
    \label{fig:micro:rtt:ltc}
\end{minipage}
\hspace{0.5cm}
\begin{minipage}[t]{0.3\textwidth}
    \centering
    \includegraphics[width=\textwidth]{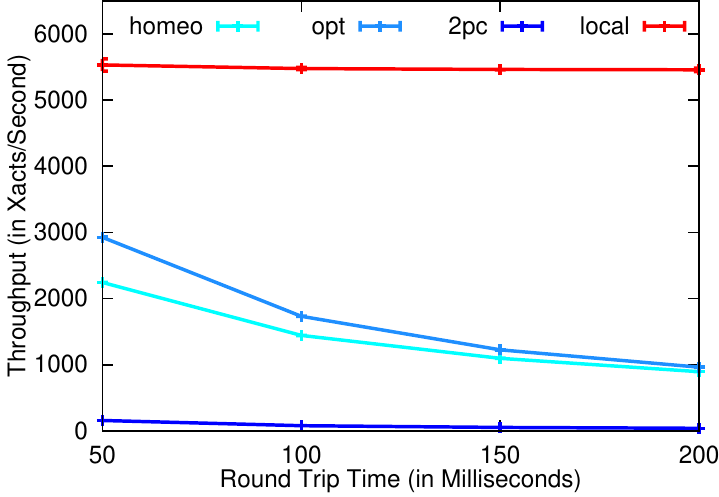}
    \caption{Throughput with network RTT ($N_r=2, N_c=16$)}
    \label{fig:micro:rtt:tps}
\end{minipage}
\hspace{0.5cm}
\begin{minipage}[t]{0.3\textwidth}
    \centering
    \includegraphics[width=\textwidth]{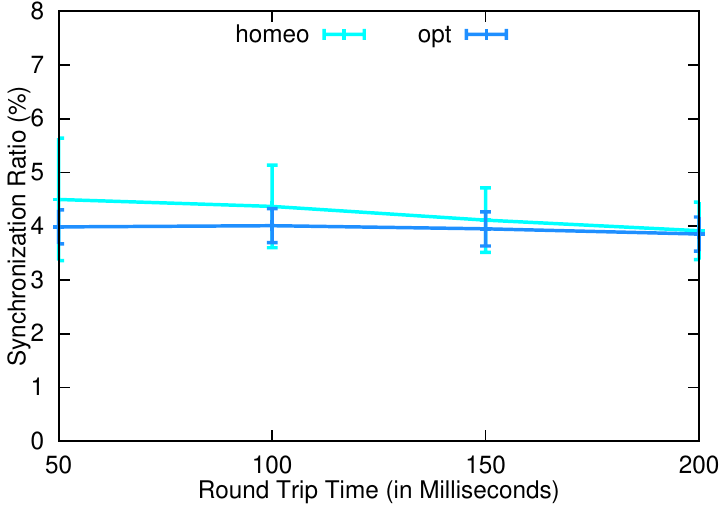}
    \caption{Synchronization Ratio with RTT ($N_r=2, N_c=16$)}
    \label{fig:micro:rtt:sync}
\end{minipage}
\end{figure*}

\begin{figure*}[ht]
\begin{minipage}[t]{0.3\textwidth}
    \centering
    \includegraphics[width=\textwidth]{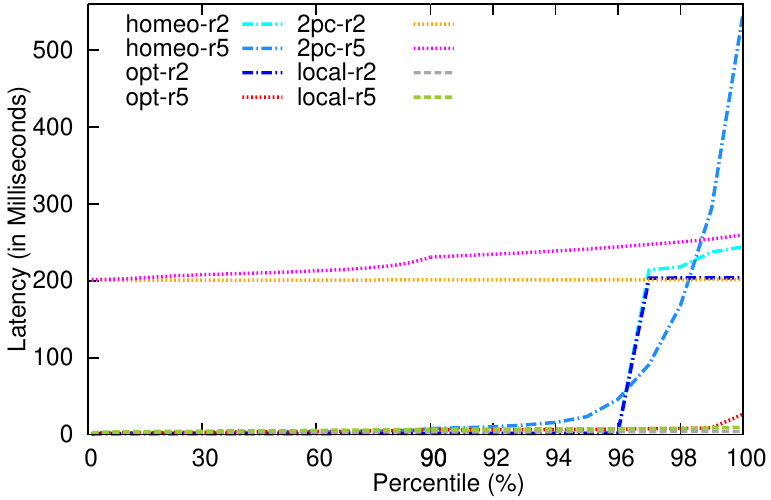}
    \caption{Latency with the number of replicas ($RTT=100ms,N_c=16$)}
    \label{fig:micro:replica:ltc}
\end{minipage}
\hspace{0.5cm}
\begin{minipage}[t]{0.3\textwidth}
    \centering
    \includegraphics[width=\textwidth]{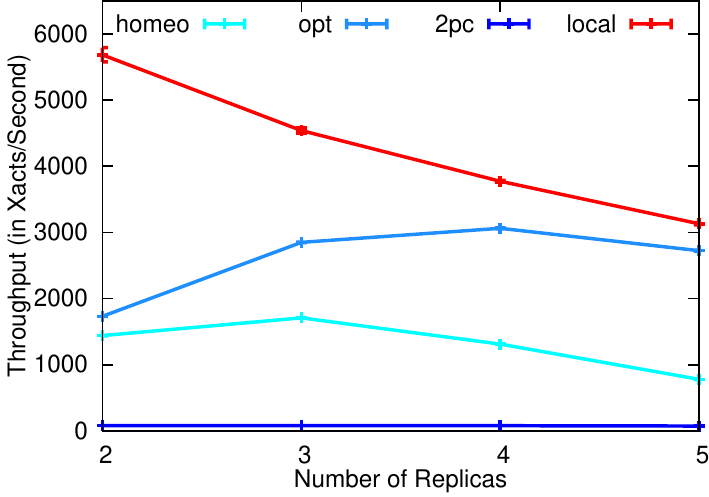}
    \caption{Throughput with the number of replicas ($RTT=100ms,N_c=16$)}
    \label{fig:micro:replica:tps}
\end{minipage}
\hspace{0.5cm}
\begin{minipage}[t]{0.3\textwidth}
    \centering
    \includegraphics[width=\textwidth]{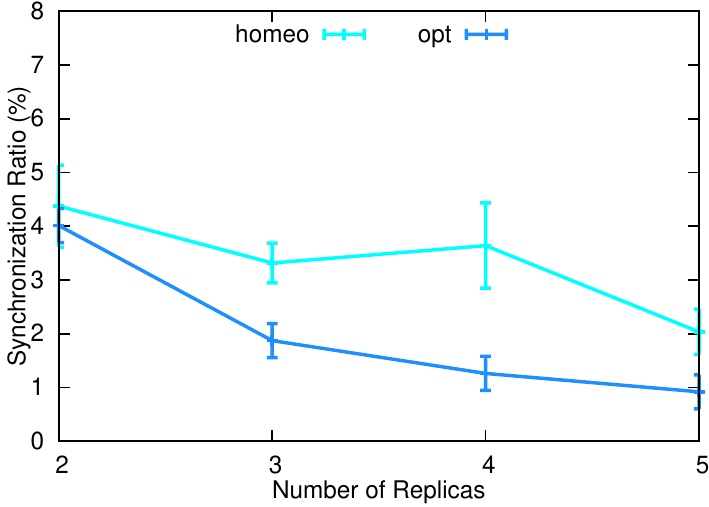}
    \caption{Synchronization Ratio with the number of replicas ($RTT=100ms,N_c=16$)}
    \label{fig:micro:replica:sync}
\end{minipage}
\vspace{-3ex}
\end{figure*}

\begin{figure*}[th]
\begin{minipage}[t]{0.3\textwidth}
    \centering
    \includegraphics[width=\textwidth]{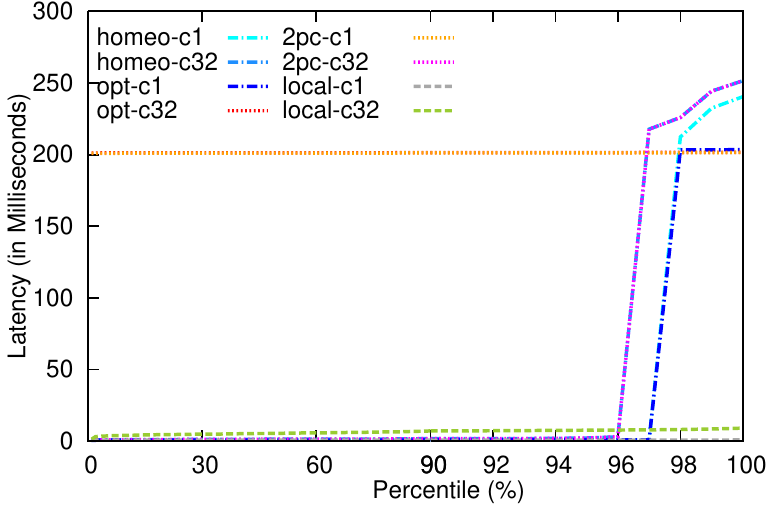}
    \caption{Latency with the number of clients ($N_r=2,RTT=100ms$)}
    \label{fig:micro:client:ltc}
\end{minipage}
\hspace{0.5cm}
\begin{minipage}[t]{0.3\textwidth}
    \centering
    \includegraphics[width=\textwidth]{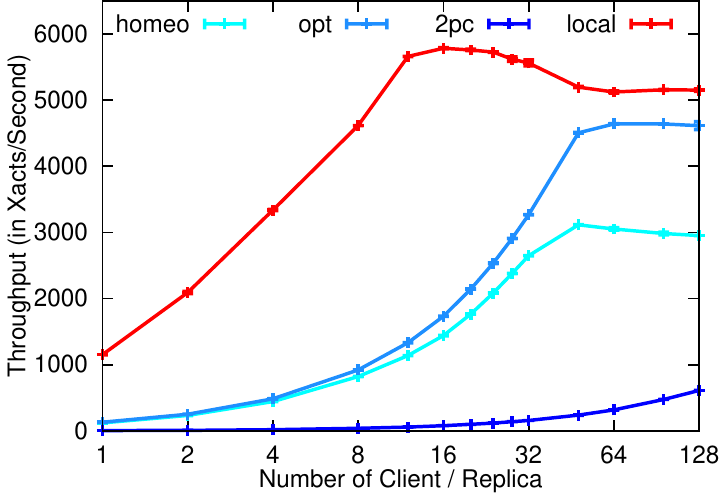}
    \caption{Throughput with the number of clients ($N_r=2, RTT=100ms$)}
    \label{fig:micro:client:tps}
\end{minipage}
\hspace{0.5cm}
\begin{minipage}[t]{0.3\textwidth}
    \centering
    \includegraphics[width=\textwidth]{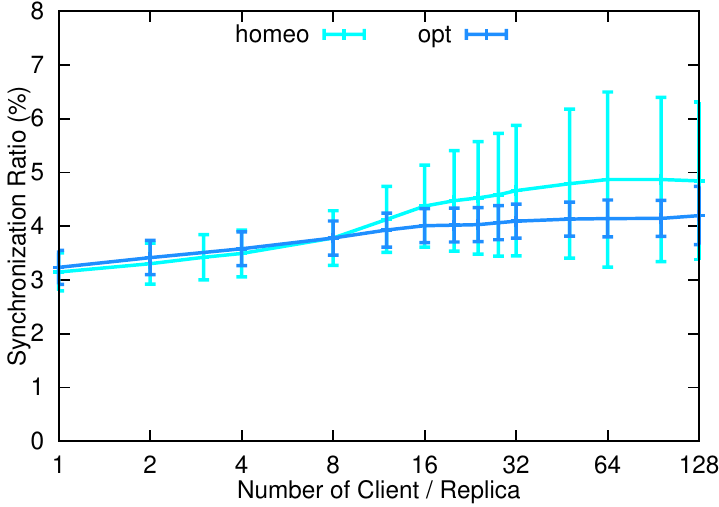}
    \caption{Synchronization Ratio with the number of clients ($N_r=2,RTT=100ms$)}
    \label{fig:micro:client:sync}
\end{minipage}
\vspace{-2ex}
\end{figure*}

\subsection{Microbenchmarks}\label{sec:micro}

With our microbenchmark, we wanted to understand how the homeostasis protocol
behaves in our intended use case -- an OLTP system where treaty violations and
negotiations are rare. In particular, we were interested in the following questions:
\begin{itemize}
\vspace{-5pt}
\item Since our protocol reduces communication, will it yield more 
performance benefits as the network round trip time (RTT) between replicas
increases?
\vspace{-5pt}
\item As the number of replicas increases, treaty negotiations will become more frequent, 
because each replica must be assigned a ``smaller''
treaty that is more likely to be violated. How does the
performance change with the degree of replication?
\vspace{-5pt}
\item Each replica/server runs multiple clients that issue transactions. How does the performance change when 
the number of clients per server (i.e., the degree of concurrency) increases?
\vspace{-5pt}
\item How much benefit can we gain from the protocol as compared to three baselines --- 
two-phase commit (2PC),  running all transactions locally without synchronizing, and a hand-crafted variant of the demarcation protocol \cite{Barbara-Milla:1994:DPT:615199.615201}?
\end{itemize}

To answer these questions, we designed a 
configurable workload inspired by an e-commerce application. 
We use a database with a single table \texttt{Stock} with just two attributes:
item ID (\texttt{itemid} INT) and quantity (\texttt{qty} INT). The item ID
is the primary key.
The workload consists of a single parameterized transaction which reads an item specified by
\texttt{itemid} and updates the quantity as though placing an order. If the 
quantity is initially greater than one, it decreases the quantity; otherwise, it refills it. 
SQL pseudocode for the transaction template is shown below.

\lstset{language=sql, frame=bt, breaklines=true, basicstyle=\ttfamily\scriptsize}
\begin{lstlisting}[caption={\small{The microbenchmark transaction; @itemid is an input parameter, while REFILL is a constant.}},
label=exp:micro:txn]
SELECT qty FROM stock WHERE itemid=@itemid;
if (qty>1) then
    new_qty=qty-1
else
    new_qty=REFILL-1
UPDATE stock SET qty=new_qty WHERE itemid=@itemid;
\end{lstlisting}

We implemented two baseline transaction execution solutions: \textit{local} and \textit{two phase commit (2PC)}. In \emph{local} mode, each replica executes the transactions locally without any communication; thus, database consistency across replicas is not guaranteed. Whereas the local mode provides a bare-bones performance baseline for how fast our transactions run locally, the 2PC mode
provides a baseline of the performance of a geo-replicated system implemented in a classical way. In addition to these baselines, we also compare against a hand-crafted solution (OPT) which exploits the transaction semantics in the same way as the demarcation protocol \cite{Barbara-Milla:1994:DPT:615199.615201}. At each synchronization point, this solution splits and allocates the remaining stock level of each item equally among the replicas. For uniform workloads, it is therefore the optimal solution. 

Our workload has several configurable parameters: network RTT, number of replicas ($N_r$),
number of clients per replica ($N_c$), and the REFILL value. By default, we set RTT to 100 ms, the number of replicas to two, the number of clients per replica to 16 and REFILL to 100. 
The database is populated with ten thousand items. 

All the experiments are run on a single Amazon EC2 c3.8xlarge instance, with 32 vCPUs,
60GB memory, and 2x320GB SSDs, running Ubuntu 14.04 and MySQL Version 5.5.38 as the local database system. 
We run all replicas on the same instance, and we simulate different RTTs. 
For each run, we start the system for 5 seconds as a warm-up phase to allow the system to reach a steady state, 
and then measure the performance for the next 300 seconds. 
All data points are averages over three runs, and error bars are given in the figures to account for the differences between runs. 

\textbf{Varying RTT} Our first experiment varies the network RTT from 50 ms to 200 ms, using the default values for all other parameters.
Figure~\ref{fig:micro:rtt:ltc} shows the transaction latency by percentile. When using the homeostasis protocol, 97\% of the transactions execute locally, with latency less than 4 ms. When a transaction requires treaty negotiation, the latency goes up to around 2RTT plus an additional overhead of less than 50 ms to find new treaties using the solver. This solver overhead manifests at the far right of Figure~\ref{fig:micro:rtt:ltc} where the latency for the homeostasis protocol is higher than for OPT for the same RTT setting.  Under 2PC, each transaction requires two RTTs, and thus the transaction latency is consistently twice the RTT. In local mode, all the transactions complete in about 2 ms.

Figure~\ref{fig:micro:rtt:tps} shows the throughput per second for each replica. 
In 2PC mode, it is less than 10 transactions per second due to the network communication cost. The homeostasis protocol allows 100x-1000x more throughput than 2PC, depending on the RTT setting. The difference between the throughput for the homeostasis protocol and local mode can be attributed to the small fraction of transactions which 
require synchronization; for example, if only 2\% of transactions require treaty negotiation and the RTT is 100ms, this leads to an average latency of 4*0.98+200*0.02=7.92ms. Finally, Figure \ref{fig:micro:rtt:sync} shows the synchronization ratio, i.e., the percentage of transactions which require synchronization under the homeostasis protocol and under OPT. The ratio is almost identical, showing that we achieve near optimal performance for this workload.

\textbf{Varying number of replicas } Next, we vary the number of replicas from 2 to 5, while setting the other parameters to their default values. 
Figure~\ref{fig:micro:replica:ltc} shows the transaction latency profile. With a higher number of replicas, 
the local treaties are expected to be more conservative and therefore lead to
more frequent violations; this leads to an increase in latency. The transaction
latency also increases for the local and 2PC modes. In the local case, this is
due to the increased resource contention since our experimental setup requires us to run all the replicas on the same server. In 2PC, each transaction stays in the system longer since it has to wait for communication with more replicas; this causes an increase in conflict rates which further increases transaction latency. 
Figure~\ref{fig:micro:replica:tps} shows the throughput per second for each
replica. As expected, the throughput decreases for all modes as the degree of replication increases. The synchronization ratio shown in Figure~\ref{fig:micro:replica:sync} also decreases with the decrease in the overall throughput of the system.

\textbf{Varying number of clients} Finally, we vary the number of clients per replica from 1 to 128, while setting the other parameters to their default values.
Figure~\ref{fig:micro:client:ltc} shows the transaction latency profile. In all modes, the transaction latency increases with the number of clients due to higher data and resource contention but is mostly dominated by network latency. 
Figure~\ref{fig:micro:client:tps} shows the throughput per replica. 
When using the homeostasis protocol with 16 clients, the throughput per client reaches 80\% of the
throughput per client we observe for 4 clients, indicating good scalability with the number of clients. 
The curve for the local mode shows a plateau or even exhibits a drop in throughput as the number of clients per replica approaches 16; with a 32-core instance, we reach a point where all cores are in use and the system is overloaded. When running the homeostasis protocol or OPT in the same case, transactions in the treaty negotiation phase free up the CPU and therefore exhibit a plateau at a higher number of clients per replica. 

We discuss additional experiments that explore the behavior of the system when we vary other parameters in Appendix~Section \ref{sec:add_exp}.

\subsection{TPC-C}\label{sec:tpcc}

\begin{figure*}[th]
\begin{minipage}[t]{0.3\textwidth}
    \centering
    \includegraphics[width=\textwidth]{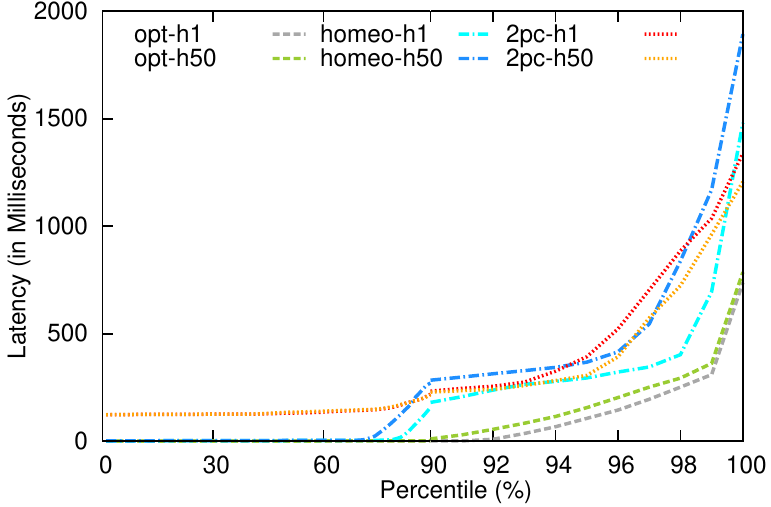}
    \caption{Latency with workload skew($N_r=2,N_c=8$)}
    \label{fig:tpcc:H:ltc}
\end{minipage}
\hspace{0.5cm}
\begin{minipage}[t]{0.3\textwidth}
    \centering
    \includegraphics[width=\textwidth]{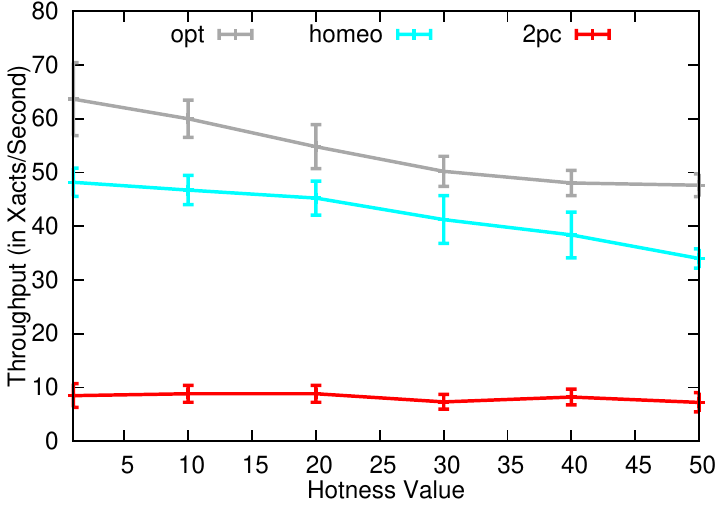}
    \caption{Throughput with workload skew($N_r=2,N_c=8$)}
    \label{fig:tpcc:H:tps}
\end{minipage}
\hspace{0.5cm}
\begin{minipage}[t]{0.3\textwidth}
    \centering
    \includegraphics[width=\textwidth]{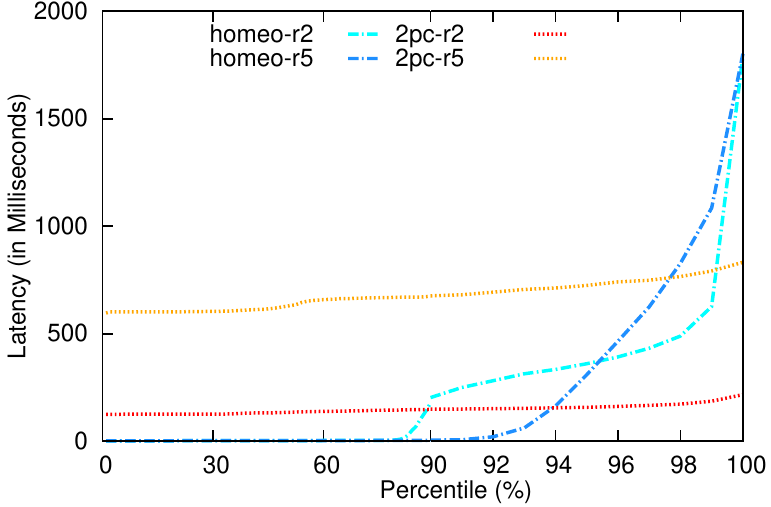}
    \caption{Latency with the number of replicas($N_c=8, H=10$)}
    \label{fig:tpcc:replica:ltc}
\end{minipage}
\vspace{-2ex}
\end{figure*}

To evaluate the performance of the homeostasis protocol over more realistic workloads with multiple transactions, larger databases, and non-uniform workload characteristics, we created a set of experiments based on the TPC-C benchmark.

\textbf{Data} The benchmark describes an order-entry and fulfillment environment. The database contains the following tables: \texttt{Warehouse}, \texttt{District}, \texttt{Orders}, \texttt{NewOrder}, \texttt{Customers},
\texttt{Items}, \texttt{Stock} and \texttt{Orderline} with attributes as specified in the TPC-C benchmark.
The initial database is populated with 10,000 customers, 10 warehouses, 10 districts per warehouse and 1000 items per district for a total of 100,000 entries in the \texttt{Stock}
table. Initial stock levels are set to a random value between 0 and 100. 

\textbf{Workload} We use three transactions based on the three most frequent transactions in TPC-C. The \texttt{New Order} transaction places a new order for some quantity (chosen uniformly at random between 1 to 5) of a particular item from a particular district and at a particular warehouse. The \texttt{Payment} transaction updates the customer, district and warehouse balances as though processing a payment; we assume that the customer is specified based on customer number. The \texttt{Delivery} transaction fulfills the oldest order at a particular warehouse and district. We explain how we encode the transactions in $\mathcal{L}$++ and what treaties are produced in the Appendix, Section \ref{sec:tpcc_encoding}.

For all experiments, we issue a mix of 45\% \texttt{New Order}, 45\% \texttt{Payment} and 10\% \texttt{Delivery} transactions.  To simulate a skew in the workload, we mark 1\% of the items as ``hot" and vary the percentage of  \texttt{New Order} transactions that order hot items. We denote this percentage as $H$. For example, a value of $H=10$ indicates that $10\%$ of all \texttt{New Order} transactions order the $1\%$ hot items. 

\textbf{Setup} We run all our experiments on c3.4xlarge Amazon EC2 instances (16
cores, 30GB memory, 2x160GB SSDs) deployed at the Virginia (UE), Oregon (UW),
Ireland (IE),  Singapore (SG) and Sao Paolo (BR) datacenters.  The average round trip
latencies between these datacenters are shown in Table~\ref{fig:rtt}.  For all
the experiments, we use a single c3.4xlarge node per datacenter. All two-replica experiments use instances from the UE and UW datacenters. There are eight clients per replica issuing transactions. All measurements are performed over a period of 500s after a warmup period of 100s. We only report measurements for the \texttt{New Order} transactions, following the TPC-C specification. For comparison, we run the same workload against an implementation of the two-phase commit protocol and a version of the homeostasis  protocol with hand crafted treaties (OPT) which minimize the expected number of treaty violations for uniform workloads. All reported values are averages of at least three runs with a standard deviation of less than 6\% in all experiments.

\begin{table}[htb]
\small
\centering
\begin{tabular}{|l|l|l|l|l|l|}\hline
 & UE & UW & IE & SG & BR\\\hline
UE & $<1$ & 64 & 80 & 243 & 164\\ \hline
UW & - & $<1$ & 170 & 210 & 227\\\hline
IE & - & - & $<1$ & 285 & 235\\\hline
SG & - & - & - & $<1$ & 372\\\hline
BR & - & - & - & - & $<1$\\\hline
\end{tabular}
\vspace{0.7em}
\caption{Average RTTs between Amazon datacenters (in milliseconds)}
\label{fig:rtt}
\end{table}

\textbf{Varying Workload Skew} For this experiment we vary $H$, i.e., the percentage of transactions that involve hot items, from 1 to 50. 
The latency profile for different values of $H$ is shown in
Figure~\ref{fig:tpcc:H:ltc}. As the value of $H$ increases, the treaties for the hot
items are violated more often, so a higher fraction of transactions takes a
latency hit. In
comparison, the latency profile for two-phase commit (2PC) is relatively
unaffected as it always incurs a two RTT latency hit. As shown in
Figure~\ref{fig:tpcc:H:tps}, the throughput for 2PC drops with
increased $H$ due to an increased rate of conflicts. The throughput for the
homeostasis protocol drops as well, but the throughput per replica is still
significantly higher than that for 2PC. Note that we only show throughput numbers for the \texttt{New Order} transaction, which constitutes 45\% of the workload. The actual number of successful transactions committed by the system per second is more than twice this value. We can increase throughput by running more clients per replica; we omit these results due to space constraints.

\begin{figure}[thb]
	\centering
    \includegraphics[width=0.28\textwidth]{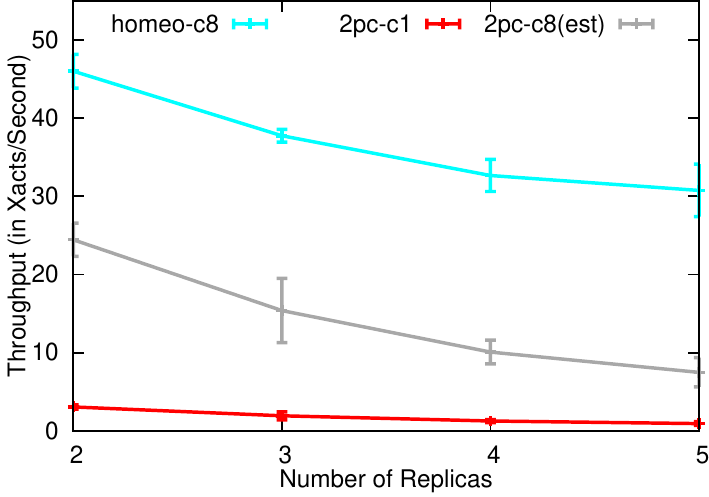}
    \caption{Throughput with the number of replicas($H=10$)}
    \label{fig:tpcc:replica:tps}
\vspace{-3mm}
\end{figure}

\textbf{Varying the number of replicas} For this experiment, we set the value of
$H$ to $10$ and measure the latency and throughput of the \texttt{New Order}
transactions as we increase the number of replicas. The replicas are added in
the order UE, UW, IE, SG and BR. The latency profile and the throughput per replica are shown in Figures~\ref{fig:tpcc:replica:ltc} and~\ref{fig:tpcc:replica:tps}. 
As we add replicas, the maximum RTT between any two replicas increases. 
This manifests itself towards the $98^{th}$ percentile as an upward shift in the latency profiles. 
With five replicas, the treaties become more expensive to compute, which also contributes to the upwards latency shift. 
On the other hand, with fewer replicas, the throughput is significantly higher, which means that with more replicas, a 
higher fraction of transactions cause treaty violations. 
This explains the leftward shift of the inflection point on the curves as the number of replicas decreases. 
In all cases, the \texttt{New Order} throughput values for the homeostasis protocol are substantially higher than the 2PC baseline. In our 2PC implementation, we only use a single client per replica: with a larger number of clients, conflicts caused frequent transaction aborts. 
Figure~\ref{fig:tpcc:replica:tps} also shows a very conservative upper bound on maximum throughput for 2PC that is obtained by multiplying the measured throughput for a single client by a factor of 8. Clearly, even this estimate still has a significantly lower throughput  than the homeostasis protocol. Thus, the homeostasis protocol clearly outperforms 2PC in all situations.

The long tail for latencies is due to the fact that the minimum allowable value of MySQL's lock wait time-out is 1 second.  In high-contention environments, more transactions time out, giving long tail latencies for high values of $H$ or high numbers of replicas. 

\textbf{Distributed Deployment} The goal of this experiment is to study the feasibility of deploying the homeostasis protocol in a realistic setting where each database replica is distributed across a number of machines. We distribute our database so that each 
machine handles all requests pertaining to one TPC-C warehouse. This distributed database is replicated across the UE and UW datacenters. We use 10 warehouses (and therefore 10 machines per datacenter), 100 districts per warehouse, 1M customers and a total of 1M entries in the stock relation. We use a transaction mix of 49\% New Order, 49\% Payment and 2\% Delivery transactions. Due to space limitations we only highlight our results here and present them more fully in the Appendix, Section \ref{subsec:appendix_tpcc_exp}.

Using the homeostasis protocol, we achieve an overall system throughput of nearly 9000 transactions per second, or around 80\% of what is achievable under OPT. As expected, the fraction of transactions requiring synchronization under homeostasis is higher than under OPT and increases as we increase the skew in the workload. Unsurprisingly, given the average latency between the two datacenters, the maximum possible throughput achievable using 2PC is around an order of magnitude smaller than with OPT.

\section{Related Work}\label{sec:relwork}

Exploiting application semantics  to improve performance in concurrent settings has been explored extensively, from thirty years ago \cite{Garcia-Molina:1983:USK:319983.319985} up to today \cite{Bailis:Coordination2014}. We refer the reader to \cite{WeikumVossen} for a more complete survey, and only highlight the approaches most similar to ours.

The demarcation protocol  \cite{Barbara-Milla:1994:DPT:615199.615201, DBLP:conf/eurosys/KraskaPFMF13} and distributed divergence control protocols \cite{Wu:1992:DCE:645477.654631,Pu:1991:RCD:115790.115856} based on epsilon serializability \cite{Ramamritham:1995:FCE:627301.627715} exploit semantics to allow non-serializable interleavings of database operations. The demarcation protocol allows asynchronous execution of transactions in a replicated system by maintaining an integrity constraint. The distributed divergence control protocol allows some inconsistency but bounds it to ensure that the schedules produced are within epsilon of serializability. More recently, Yu and Vahdat proposed a continuous consistency model for replicated services \cite{Yu:2000:DEC:1251229.1251250} which allows bounded inconsistency and uses anti-entropy algorithms \cite{Olston:2001:APS:375663.375710, Olston:2000:OPT:645926.671877} to keep it at an acceptable level. As discussed in Section \ref{subsec:scope}, the distinguishing feature of our work is the use of program analysis techniques to automatically infer the constraints to be maintained, and the fact that the homeostasis protocol is independent both of the transaction language and the constraint language.

The Calvin \cite{Thomson:2012:CFD:2213836.2213838} and Lynx \cite{Zhang:2013:TCA:2517349.2522729} systems also perform static analysis of transactions to improve performance; however, they ensure that all transactions operate on consistent data, whereas we allow them to see inconsistent data if the inconsistency is ``irrelevant.''

There is extensive research on identifying cases where transactions or operations commute and allowing interleavings which are non-serializable yet correct \cite{Kumar:1988:SBT:50202.50215, Li:2012:MGS:2387880.2387906, conf/ppopp/HerlihyK08, CRDT, DBLP:conf/cidr/AlvaroCHM11, 183989}. Our solution  extends this work, as the use of LR-slices allows transactions to commute in special cases where they may not commute in general. The approach in \cite{Clements:2013:SCR:2517349.2522712} allows for state-dependent commutativity, but at the interface level rather than the read/write level. $\mathcal{I}$-confluence invariants \cite{Bailis:Coordination2014} and warranties  \cite{Liu:2014:WFS:2616448.2616495} both have similarities to our treaties, but neither is inferred automatically from code.

Some systems allow inconsistency that does impact transaction semantics when deemed acceptable cost-wise for the application \cite{Kraska:2009:CRC:1687627.1687657, Bailis:2012:PBS:2212351.2212359}.  Including the cost of inconsistency into our treaty optimization metric is future work.

There is more general distributed systems work that explains how to detect whether global properties hold \cite{Chandy:1985:DSD:214451.214456,Marzullo91detectionof}, and how to enforce them using local assertions \cite{Carvalho:1982:DA:800220.806689}. However, this work does not address how the local assertions may be computed. 

\ifconferenceversion
\else 
Symbolic tables bear similarities to constructs from Owicki-Gries proofs \cite{susan1976axiomatic} used in reasoning about correctness of parallel programs; however, we are applying them in a very different setting.
\fi

There is a large body of work on using synthesis to develop concurrent programs \cite{cherem2008inferring, vechev2010abstraction, solar2008sketching, hawkins2012concurrent}. Our applications are database transactions rather than general-purpose code, and we deal with extra challenges such as the need to optimize based on workload.

\section{Conclusion}\label{sec:conclusion}

We have presented the homeostasis protocol, which allows transactions to execute correctly  in a multi-node system while minimizing communication. We extract consistency requirements from transaction code \emph{automatically via program analysis} rather than requiring human input, which enables a fully automated adaptive consistency management protocol. We have presented the method as a general framework and provided concrete implementations for each component in the framework.

\section{Acknowledgments}

We would like to thank Peter Bailis and the anonymous SIGMOD reviewers for their insightful comments.

This research has been supported by the NSF under Grants CNS-1413972,
CCF-1253165, CNS-1012593 and IIS-0911036, by the European Research Council under grant ERC Grant 279804, by the iAd Project funded by the Research Council of Norway and by a Google Research Award. Any opinions, findings, conclusions or recommendations
expressed in this paper are those of the authors and do not
necessarily reflect the views of the sponsors.

\small
\bibliographystyle{abbrv}
\bibliography{references}

\normalsize
\newpage
\appendix
\section{Expressing transactions in $\mathcal{L}$}\label{subsec:expressiveness}

In this section, we explain how to translate programs in a higher-level language such as SQL or $\mathcal{L}$++ to $\mathcal{L}$. The translation lets us formalize the intuition that adding native support for bounded relations to $\mathcal{L}$ does not alter the expressive power of the language.

$\mathcal{L}$ does not have built-in support for relations or arrays. However, these data structures and read/write accesses to them can be simulated using nested if-then-else statements. An array $a$ of length $n$ is stored in the database as a set of objects $\{a_0, a_1, \ldots, a_{n-1}\}$. The following code reads the $\hat{i}$th element of the array $a$ into a temporary variable $\hat{x}$:
\begin{align*}
                 & \texttt{if } \hat{i} = 0 \texttt{ then } \hat{x} := \texttt{read}(a_0)\\
  \texttt{else } & \texttt{if } \hat{i} = 1 \texttt{ then } \hat{x} := \texttt{read}(a_1)\\
  \texttt{else } & \texttt{if } \ldots \texttt{ else } \texttt{skip}
\end{align*}
We use the command $\hat{x} := \texttt{read}(a(\hat{i}))$ as syntactic sugar for the code above. Analogously, we use the command $\texttt{write}(a(\hat{i}) := \hat{x})$ as syntactic sugar for a sequence of if-then-else statements that writes the value of the temporary variable $\hat{x}$ into the $\hat{i}$th element of $a$.

Once we have one-dimensional arrays, we can extend them to two-dimensional arrays in the standard manner, by storing a two-dimensional array in row-major order inside a single-dimensional array and accessing elements by offset computation. Thus $a(\hat{i}) (\hat{j})$ (accessing the $i, j$th element in the two dimensional array $a$) is syntactic sugar for $a(\hat{i}*D + \hat{j})$, where $D$ is the size of the array along the first dimension.

Two-dimensional arrays allow us to represent relations (tables). If we  assume a bound on the maximum number of tuples in any relation, we can encode bounded (foreach) iteration over relations and express \texttt{SELECT-FROM-WHERE} clauses as a sequential scan over the entire relation. In the sequential scan, for each tuple in the array we check whether it matches the selection condition using if-then-else statements.

For SQL \texttt{UPDATE} statements, database insertions and deletes are modeled by ``preallocating'' extra space in the array and keeping track of used vs. unused space with suitable placeholder values for unused portions of the array.

The above encoding is very straightforward and can easily be performed automatically by a compiler. Encoding more expressive language constructs than \texttt{SELECT--FROM--WHERE} will require more elaborate compilation techniques. Depending on the source language, the process may run into the theoretical expressiveness limits of $\mathcal{L}$; however, we have determined (through manual inspection) that $\mathcal{L}$ is expressive enough to encode all five TPC-C transactions, which are representative of many realistic OLTP workloads.

It is possible to add extensions to $\mathcal{L}$ and make it more expressive so it  provides support for unbounded iteration or complex data structures. These extensions are compatible with our analysis, as long as we add appropriate rules for symbolic table computation and understand that the symbolic table formulas $\varphi$ may need a more expressive logical language.

\section{Handling remote writes}\label{subsec:remotewrites}

In this section, we explain how to adapt the homeostasis protocol to handle non-local writes, removing the need for Assumption \ref{assum:local}. The most common use case for  remote writes is replicated systems, where -- conceptually -- a write to an object requires a remote write on each site which keeps a replica of the object. 

The basic idea is to add extra logic to the existing protocol that transforms arbitrary transactions into transactions containing only local writes prior to execution. The transformations are conceptually simple and are easily applied automatically, without human intervention. We first explain in detail how to do this for programs in our language $\mathcal{L}$ from Section \ref{subsec:sem_tables}, and then discuss how the transformation would apply to programs in other and/or richer languages.

Given a set of transactions in $\mathcal{L}$, we proceed as follows. For each database object $x$ we introduce a set of fresh database objects $\{ dx_1, dx_2, \ldots, dx_M \}$ -- one for each site that runs transactions that perform remote writes on $x$. Each $dx_i$ is local to site $i$ and is initialized to $0$. We modify any transaction executed on site $i$ so it writes to $dx_i$ instead of $x$.

An example of this transformation is shown in Figure \ref{fig:delta-transformation}. We assume that the original transaction (Figure \ref{fig:delta-transformation:original}) runs on site 1, but $x$ resides on some other site. To keep the example simple, we also assume that this transaction is the \emph{only} transaction in the system that reads or writes to $x$.

The transformed transaction (Figure \ref{fig:delta-transformation:transformed}) writes to the local  $dx_1$ instead of the remote $x$; however, it maintains the invariant that $x + dx_1$ will always be equal to the ``real'' value of $x$. Thus every remote write is replaced with a local one. However, synchronization may still be required because of remote reads. For example, the transaction from Figure \ref{fig:delta-transformation:transformed} must always perform a remote read to determine the  value of $x$.

\begin{figure}[!t]
 \begin{subfigure}[b]{0.24\textwidth}
   \begin{tabular}{l}
      $\hat{x} \coloneqq \texttt{r}(x)$\\
      $\texttt{if } 0 < \hat{x} \texttt{ then }$\\
      $~~~ \texttt{w}(x = \hat{x} - 1)$\\
      $\texttt{else}$\\
      $~~~ \texttt{w}(x = 10)$\\
   \end{tabular}
   \caption{Original transaction}
   \label{fig:delta-transformation:original}
 \end{subfigure}
 ~
 \begin{subfigure}[b]{0.24\textwidth}
   \begin{tabular}{l}
      $\hat{x} \coloneqq \texttt{r}(x) + \texttt{r}(dx_1)$ \\
      $\texttt{if } 0 < \hat{x} \texttt{ then }$\\
      $~~~ \texttt{w}(dx_1 = \hat{x} - 1 - \texttt{r}(x))$\\
      $\texttt{else}$\\
      $~~~ \texttt{w}(dx_1 = 10 - \texttt{r}(x))$\\
   \end{tabular}
   \caption{Transformed for Site 1}
   \label{fig:delta-transformation:transformed}
 \end{subfigure}
 
\vspace{2ex}
 \begin{subfigure}[b]{0.48\textwidth}
   \begin{tabular}{l}
     $\texttt{if } 0 < \texttt{r}(x) + \texttt{r}(dx_1) \texttt{ then }$\\
     $~~~ \texttt{w}(dx_1 = \texttt{r}(dx_1) - 1)$\\
     $\texttt{else}$\\
     $~~~ \texttt{w}(dx_1 = 10 - \texttt{r}(x))$\\
   \end{tabular}
   \caption{Transformed and simplified}
   \label{fig:delta-transformation:simplified}
 \end{subfigure}
 
 \caption{Transforming transactions to eliminate non-local writes. The \texttt{read} and \texttt{write} commands are abbreviated as \texttt{r} and \texttt{w} respectively.}
 \label{fig:delta-transformation}
\end{figure}

If we had other transactions in the system that read $x$, they would need to be transformed to read both $x$ and $dx_1$ each time, and to work with the sum of these two values instead of just the value of $x$. If we had transactions running on sites other than $i$ that \emph{write} to $x$, we would need to introduce additional $dx_i$ objects and apply appropriate transformations. In the common case where $x$ is a replicated object that is read and written on each of $K$ sites, we would introduce a $dx_i$ for every site. Every statement of the form $\texttt{read}(x)$ would be translated to 
$$\texttt{read}(x) + \sum_{j~=~1~\text{to}~K}~ \texttt{read}(dx_j)$$
Every $\texttt{write}(x = \hat{x})$ for a transaction running on site $i$ would be translated to 
$$\texttt{write}(dx_i) = \hat{x} - \texttt{read}(x) - \sum_{j~=~1~\text{to}~K, ~ j \neq i} ~\texttt{read}(dx_j)$$

The next step is to eliminate as many remote reads as possible using semantics-preserving program transformations. In our running example, the resulting transaction for Site 1 is shown in Figure \ref{fig:delta-transformation:simplified}. This transaction replaces the expression $\hat{x} - 1 - \texttt{r}(x)$, which references the remote variable $x$, with the equivalent expression
\begin{equation*}
  (\texttt{r}(x) + \texttt{r}(dx_1)) - 1 - \texttt{r}(x)
  \quad=\quad \texttt{r}(dx_1) - 1
\end{equation*}
which does not. This simplified transaction can avoid remote reads much of the time: if we can generate a treaty which ensures that $0 < x + dx_1$ then the transaction can be executed on Site 1 using only locally available information about the value of $dx_1$.

All the transformations described above are easy to perform automatically. Once we have applied the transformations to all transaction code, we can run the system using the homeostasis protocol as described previously, because Assumption \ref{assum:local} now holds. In practice, we might initialize the $dx$ objects to $0$ and reset them to $0$ at the end of each protocol round, but this is not required for correctness. The protocol still yields provably correct execution; however, to complete the proof we now need to show that the above transformations preserve the observable behavior of transactions. A full treatment of the issue requires us to define formal evaluation semantics for $\mathcal{L}$, and is beyond the scope of this paper.

For transactions in a language that is richer than $\mathcal{L}$ and may support data types other than integers, it is still possible to perform the above transformation if each data type comes with a suitable \emph{merge function} that is the equivalent of addition in the integer case. Under this merge function, elements of the data type must form an Abelian group. For example, we can apply the transformation to multisets (bags) of integers, using bag union as the merge function. 

If the data type we wish to use does not come with a suitable merge function, the above  transformations cannot be used and it is necessary to synchronize on every update to a replicated object of this type. However, existing research has shown that there are numerous  commutative replicated data types which do come with the merge functions we need \cite{CRDT}. Thus we expect our transformation technique to be broadly applicable in practice.

\section{Treaty computation details}\label{sec:localmisc}

\subsection{Preprocessing the formula $\psi$}

The preprocessing of $\psi$ involves identifying all subexpressions $\theta$ that prevent it from being a conjunction of linear constraints, for example $\theta$ may be a negated boolean expression $\neg b$. For each such expression, we replace it within $\psi$  by the actual value of $\theta$ on $D$. In addition, if the variables in $\theta$ are ${x_1, x_2, \cdots   x_n}$, we transform $\psi$ to $\psi \wedge \bigwedge_i~(x_i = D(x_i))$. Intuitively, any variables involved in the subexpression have their values fixed to the current ones. It is clear that the transformed $\psi$ implies the original one so we can use it in what follows without sacrificing correctness.

\subsection{Finding a good treaty configuration}

Algorithm \ref{alg:goodwarranty} computes valid treaty configurations. It has two tunable parameters: a lookahead interval $L$ and a cost factor $f$. We examine $f$ possible future system executions, where each execution is a sequence of $L$ transactions and is constructed using our workload model. We search for a configuration that minimizes the number of treaty violations in our $f$ executions. For larger $f$ and $L$, the probability of a future treaty violation decreases, but the search for an optimal configuration takes more time to run.

The configuration search is performed by a suitable MaxSAT solver. We assume that we have an efficient solver that takes as input a first-order logic formula, finds the largest satisfiable subset of constraints that includes all the \emph{hard constraints} and also produces a model (a concrete assignment of values to configuration variables) which satisfies this subset of constraints.                                                   

We now discuss the algorithm in more detail. The algorithm first generates $\theta_h$---this is a formula that says the local treaties must enforce the global treaty. This must always be true for a valid treaty configuration, so we will have $\theta_h$ as a hard constraint (Line 4). Next (Lines 6-12) the algorithm generates $f$ sequences of transactions of length $L$ using the workload model. For each sequence, it simulates the execution of transactions in the sequence to produce a sequence of databases, one per transactional write (Line 8). The desired behavior is that none of the databases in the sequence violate the local treaties. We therefore add an appropriate constraint for each database in the sequence (Line 10). This process yields a set $\Theta_s$ of soft constraints, which are fed to the MaxSAT solver together with the hard constraints in $\theta_h$ (Line 13).

\begin{algorithm}
 \begin{algorithmic}[1]
  \STATE Let $L$ be the lookahead interval.
  \STATE Let $f$ be the cost factor.
  \STATE Let $\varphi = \bigwedge_{i \in [1, K]} \varphi_{\Gamma_i}$ \COMMENT {Conjunction of local treaty templates.} 
  \STATE $\theta_h = \{\forall~ Vars(\varphi_\Gamma),~ \bigwedge_{\{i \in [1:K]\}}(\varphi_{\Gamma_i}) \Rightarrow
\varphi_{\Gamma}\}$ \COMMENT {Local treaties enforce global treaty.}
  \STATE $\Theta_s = \emptyset$
  \FOR {$i = 1 \text{ to } f$}
  \STATE Generate a sequence $S$ of $L$ transactions using model
  \STATE Let $[D_0, D_1, ..., D_L]$ be the sequence of databases obtained by executing $S$ on $D$ (we have $D_0 = D$).
  \FOR{$j = 1 \text{ to } L$}
  \STATE $\Theta_s = \Theta_s \cup \{\varphi(D_j)\}$ \COMMENT {Treaty holds on $D_j$.}
  \ENDFOR
  \ENDFOR
  \STATE $C = MaxSAT(\{\theta_h\} \cup \Theta_s)$ \COMMENT{Hard constraints $\theta_h$, soft $\Theta_s$}
   \end{algorithmic}
 \caption{Finding a valid treaty configuration that reduces the probability of future violation.}
 \label{alg:goodwarranty}
\end{algorithm}

We illustrate an execution of Algorithm \ref{alg:goodwarranty} by continuing with our example and our local treaty templates $\varphi_{\Gamma_1} :  (x + c_y \geq 20)$ and
$\varphi_{\Gamma_2} :  (c_x + y \geq 20)$. We have $\theta_h = (\forall [x, y], \varphi_{\Gamma_1} \wedge \varphi_{\Gamma_2} \Rightarrow \varphi_\Gamma)$.

Suppose that $L = 3$ and $f = 3$ and transaction $T_1$ is twice as likely as $T_2$. We construct $3$ possible transaction sequences of length $L$ drawn from this distribution. Suppose these are $S_1 = [T_1; T_1; T_2]$, $S_2 = [T_1; T_1; T_1]$ and $S_3 = [T_1; T_2; T_1]$. If we execute $S_1$ on the initial database we obtain the sequence of states $[(10, 13); (9, 13); (8, 13); (8, 12)]$ where each element represents an ordered pair of $x$ and $y$ values. This sequence can be computed by looking up in the symbolic table the partially evaluated transaction for each transaction in $S_1$ and applying it.

The desired behavior is that no database in the above sequence violates a local treaty. Thus $\varphi_{\Gamma_1} \wedge \varphi_{\Gamma_2}$ must hold on each of the four databases. Plugging in the $x$ and $y$ values from each database into $\varphi_{\Gamma_1} \wedge \varphi_{\Gamma_2}$, taking the conjunction and performing some simplification yields the soft constraints $\{(c_y \geq 12), (c_x \geq 8)\}$. Repeating the procedure for $S_2$ and $S_3$, we get soft constraints $\{(c_y \geq 13), (c_x \geq 7)\}$ and $\{(c_y \geq 12), (c_x \geq 8)\}$.

By passing all these constraints to a MaxSAT solver, we find that not all of the soft constraints can be satisfied in addition to $\theta_h$.
However, for $c_y = 12$ and $c_x = 8$, we can satisfy both the first and third set of soft constraints. The algorithm therefore chooses
$C = \{(c_y = 12, c_x = 8)\}$. Note that this configuration allows more flexibility to the site at which $T_1$ runs, since it is more frequent. As desired, we have minimized the probability
of a treaty violation subject to the computational constraints $f$ and $L$.

\subsection{Lifting Assumption 4.1}

We now explain how to handle transactions that violate Assumption \ref{assum:noremoteread}. An example of such a transaction would be one that reads the value of remote object $y$ and copies it into the local object $x$. Intuitively, if we wish to avoid the remote read of $y$ yet guarantee correct execution, we need a global treaty that guarantees the value of $y$ will never change. We can achieve this by adding a constraint to the local treaty \emph{on the site where $y$ resides}. To do this, we need to perform a post-processing step on the local treaty templates. Specifically, for every transaction $T$ executing at site $i$, we consider the partially evaluated transaction $\phi_T$ found in the relevant row of the symbolic table. For every variable $x$ such that $\texttt{read}(x)$ appears in $\phi_T$ but $Loc(x) \neq i$, we substitute it with a new local variable $c_x$ and modify $\varphi_{\Gamma_{Loc(x)}}$ to $\varphi_{\Gamma_{Loc(x)}} \wedge (x = c_x)$. Once this postprocessing step is complete, we can run Algorithm \ref{alg:goodwarranty} as before.

\section{Examples beyond top-$k$}\label{sec:linear_and_nonlinear}

In this section, we elaborate on the examples from in Section \ref{subsec:scope}.

The first example is the ``maximum of minimums'' case, which we encode as a program which maintains a list of lists in a two-dimensional array. In our weather scenario, each list represents a day and contains the temperature measurements for that day. The program adds a value to a specific list, finds the minimum value in each list, finds the top $k$ values among these minimums and prints them  to the screen.

The output of the program changes if the inserted value is a minimum in its own list and is one of the $k$ highest minimum values among the other lists. Thus, there are $k + 2$ high-level cases to handle: one where the new value is not a minimum in its own list and the remaining $k + 1$ for the case where it is, and has a particular relative ordering with respect to the current top-$k$ minimums. For each of these cases, the program will print a different result. 

The actual local treaties required for correct disconnected execution at each site depend on the above information as well on what (portions of the) lists are stored on each site. For instance, if each list is stored on a different site, it is necessary to synchronize  when a new minimum value is added to the local list, and this value is higher than the lowest value in the current top-$k$. If multiple lists are stored on one site or a list is distributed (or indeed replicated) across multiple sites, the treaties will be more complex. The treaties can be expressed as conjunctions of linear constraints and can in principle be derived manually, although the derivation process will require some care if the lists are stored on the various sites according to one of the more complex schemes we described.

Our second example looks for the top-$k$ days with the highest temperature difference, where the temperature difference is the daily high temperature minus the daily low temperature. It can be implemented like the previous example, but now upon inserting a value we find and print the indexes of the top $k$ lists which have the highest difference between max and min. 

There are two high-level cases that affect how this program behaves -- either the list that was updated enters the top-$k$ after the update or it does not. If it does not, there are three possible reasons and subcases. Either the the new value that is inserted is neither a new min nor a new max in its list,  or it is a new max but the resulting new min-max difference for that list is not greater than the current top-$k$, or it is a new min but the resulting new min-max difference is not greater than the current top-$k$. If the updated list does enter the top-$k$, we need $k+1$ subcases for each relative ordering. As before, the actual treaties required for correct disconnected execution will depend on what is stored where.

Like the first example, this too can in principle be handled manually by a careful and motivated human treaty generator. However, it is unclear how much more complexity can be added without overwhelming the human and introducing errors. On the other hand, our analysis can compute correct symbolic tables and local treaties for both examples automatically.

\section{Analyzing TPC-C transactions}\label{sec:tpcc_encoding}

In this section, we explain how to express the TPC-C transactions  \texttt{New Order}, \texttt{Delivery} and \texttt{Payment} in $\mathcal{L}$++, and we explain what treaties the homeostasis protocol produces for them.

\subsection{Encoding the TPC-C transactions}

Recall that the \texttt{New Order} transaction places a new order for some quantity (between 1 to 5) of a particular item from a particular district and at a particular warehouse.  The \texttt{Payment} transaction updates the customer, district and warehouse balances as though processing a payment, and the \texttt{Delivery} transaction fulfills orders at a particular warehouse and district. 

Expressing these three transactions in $\mathcal{L}$++ is largely straightforward, since they consist primarily of a sequence of SQL \texttt{SELECT}, \texttt{UPDATE}, \texttt{INSERT} and \texttt{DELETE} statements, and map directly to relation read and write statements in $\mathcal{L}$++ that are syntactic sugar for $\mathcal{L}$ code, as discussed in the Appendix Section \ref{subsec:expressiveness}. 

\texttt{New Order} and \texttt{Delivery} both write to the \texttt{NEW-ORDER} table; the TPC-C specification requires an ordering where \texttt{New Order} creates orders with monotonically increasing order ids for a given warehouse and district, whereas \texttt{Delivery} fulfills the oldest order (i.e. the one with the smallest id) for a given warehouse and district each time. However, the specification does not describe how this ordering should function in a replicated version of TPC-C and what constraints it needs to obey.

For our implementation, we assume that the ordering must exist and that \texttt{Delivery} must fulfill the lowest-numbered order each time, but that the assignment of new order ids does not need to be monotonic across all sites. That is, each site generates monotonically increasing order ids (for each district and warehouse) and no two sites can ever generate the same order id. Thus we ensure the existence of a global ordering, but no synchronization is required when \texttt{New Order} executes.

Finally, to make our code suitable for a replicated setting, we apply transformations as described in Appendix Section \ref{subsec:remotewrites}. 

\subsection{Treaties for TPC-C}

\begin{figure*}
\begin{minipage}[t]{0.3\textwidth}
    \centering
    \includegraphics[width=\textwidth]{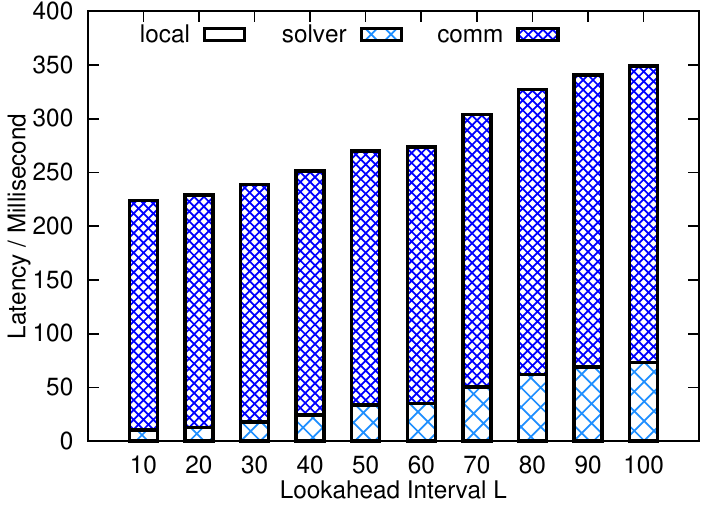}
    \caption{Transaction Latency Breakdown with Lookahead Interval L ($RTT=100ms, N_c=16, REFILL=100)$}
    \label{fig:micro:refill:bk}
\end{minipage}
\hspace{0.5cm}
\begin{minipage}[t]{0.3\textwidth}
    \centering
    \includegraphics[width=\textwidth]{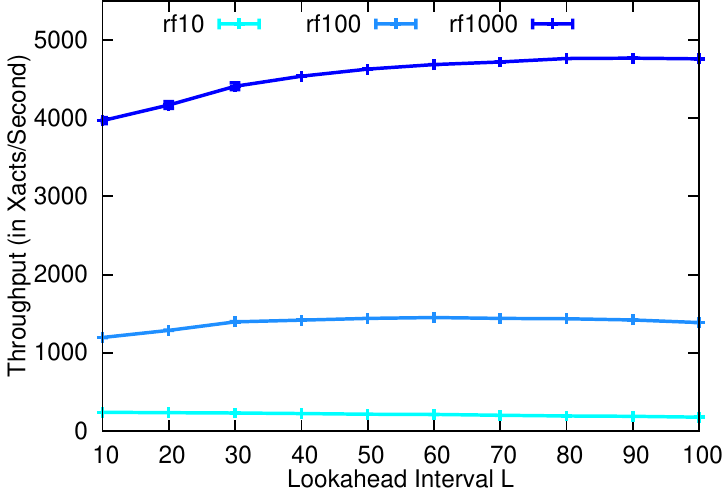}
    \caption{Throughput with Lookahead Interval L for different REFILL values ($RTT=100ms,N_c=16$)}
    \label{fig:micro:refill:tps}
\end{minipage}
\hspace{0.5cm}
\begin{minipage}[t]{0.3\textwidth}
    \centering
    \includegraphics[width=\textwidth]{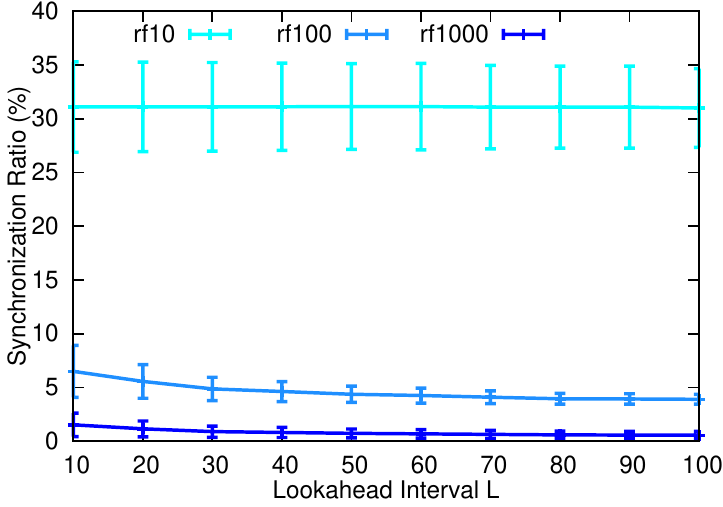}
    \caption{Synchronization Ratio with Lookahead Interval L for different REFILL values ($N_r=2,RTT=100ms,N_c=16$)}
    \label{fig:micro:refill:sync}
\end{minipage}
\end{figure*}

\begin{figure*}
\begin{minipage}[t]{0.3\textwidth}
    \centering
    \includegraphics[width=\textwidth]{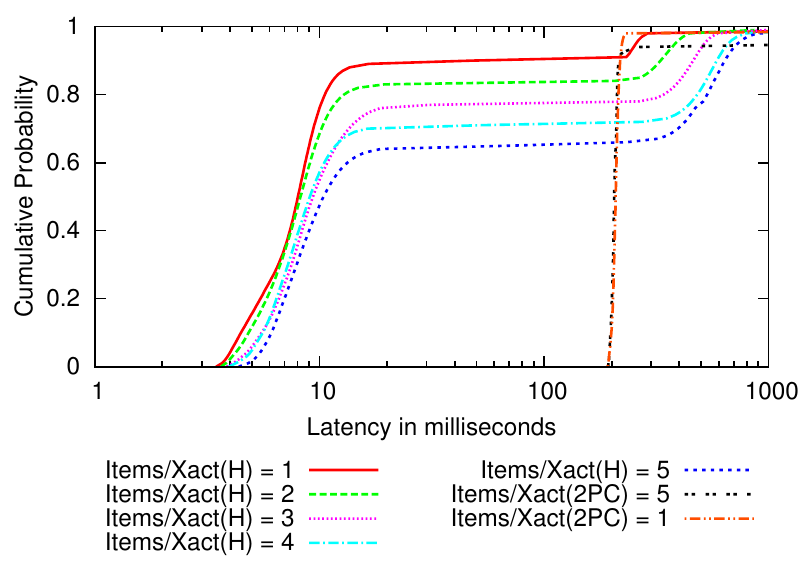}
    \caption{Latency with number of items accessed per transaction ($RTT=100, REFILL=100, N_c=20, L=20$)}
    \label{fig:microbm:item:ltc}
\end{minipage}
\hspace{0.5cm}
\begin{minipage}[t]{0.3\textwidth}
    \centering
    \includegraphics[width=\textwidth]{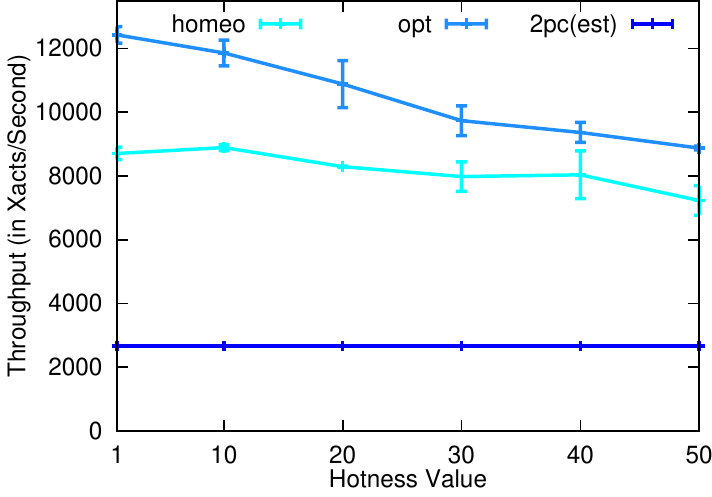}
    \caption{Overall system throughput with H ($REFILL=100,N_c=16$)}
    \label{fig:tpcc:exp1_new:tps}
\end{minipage}
\hspace{0.5cm}
\begin{minipage}[t]{0.3\textwidth}
    \centering
    \includegraphics[width=\textwidth]{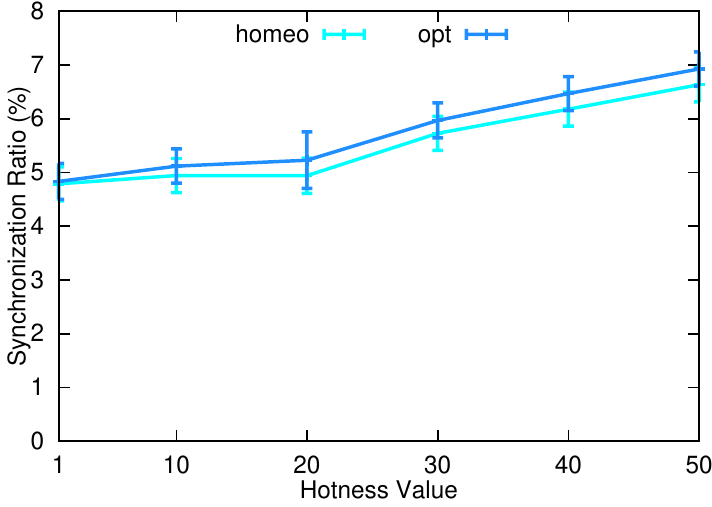}
    \caption{Synchronization Ratio with H ($REFILL=100, N_c=16$)}
    \label{fig:tpcc:exp1_neq:sync}
\end{minipage}
\end{figure*}

We now explain in more detail what treaties were generated from our $\mathcal{L}$++ implementation described above. 

The \texttt{Payment} transaction with customer specified by number updates several  balance amounts. These amounts are never read by any transaction, including other instances of \texttt{Payment}. As explained in Appendix Section \ref{subsec:remotewrites}, incrementing a balance by a given amount can be done without reading the amount itself if we use auxiliary local objects to store the local ``deltas'' of the balance. Thus, no treaties are required on the balance values for the homeostasis protocol, and instances of \texttt{Payment} run without ever needing to synchronize.

The \texttt{New Order} transaction requires treaties on the stock values for each item and for each warehouse, i.e. on the \texttt{S\_QUANTITY} for each row of the \texttt{STOCK} table in the TPC-C schema. The global treaty for each such quantity $q$ ensures that $q$ may never be negative, i.e. $q >= 0$. The local treaties generated by our protocol constrain $q$ further and have the form $dq_i > c$, where $dq_i$ is the auxiliary local object corresponding to $q$ on site $i$ (see Appendix Section \ref{subsec:remotewrites}) and $c$ is chosen through optimization as explained in Section \ref{sec:treaties}.

The \texttt{Delivery} transaction, as explained above, needs to delete a tuple from  the \texttt{NEW--ORDER} table which has the lowest order id for a given district and warehouse. This does require synchronization. In our $\mathcal{L}$++ code, we keep a database object for each warehouse and district that stores the current lowest order id in \texttt{NEW--ORDER} for that warehouse and district. Each such object has an associated global treaty that fixes it to its current value in the database. Because \texttt{Delivery} also updates (increments) that object, executing it causes a treaty violation and forces a synchronization. The only transaction reading or writing to that object is \texttt{Delivery}, and so synchronization occurs every time that transaction runs. 

In addition, we need to ensure that the \texttt{Delivery} transaction never sees an empty \texttt{NEW--ORDER} table unless the table is truly empty (rather than containing inserts which have not been propagated to the site running \texttt{Delivery}). To achieve this, we maintain a treaty for each warehouse and district that requires the number of unfulfilled orders to remain above zero. If a \texttt{Delivery} transaction processes the order with the \emph{highest} order id that it is ``aware of,'' this will cause a violation of the treaty and synchronization will occur.

In summary, for the TPC-C transactions we use, there is a global treaty of the form $q > 0 $ for every quantity $q$ in the \texttt{STOCK} table. There is also a treaty for the current lowest unprocessed order id in the \texttt{NEW-ORDER} table for each warehouse and district, which fixes that order id to its current value. Finally, there is a treaty for each warehouse and district stating that the highest unprocessed order id is greater than the lowest unprocessed order id. 

\vspace{-0.75em}
\section{Additional Experiments}\label{sec:add_exp}

In this section, we present a number of additional experiments.

\subsection{Microbenchmark setting}

\textbf{Varying the lookahead interval} Within the microbenchmark setting from Section \ref{sec:micro}, we experiment with using the homeostasis protocol and varying the lookahead interval $L$ used in Algorithm \ref{alg:goodwarranty} from Appendix Section \ref{sec:localmisc}. Varying this parameter trades off the time overhead of treaty generation versus the quality of the generated treaties, where a better treaty is one that requires less frequent synchronization. We also vary the value of REFILL, since a larger REFILL value allows for more flexible treaties. The REFILL values we used were 10, 100 and 1000.

Figure~\ref{fig:micro:refill:bk} gives the time breakdown for transaction latencies of transactions which caused a treaty violation as we vary $L$, showing time spent in actual transaction execution (\emph{local}), in communication with other replicas(\emph{comm}), treaty computation and in generating treaties using Algorithm \ref{alg:goodwarranty} (\emph{solver}). The values for \emph{local} are negligibly small as compared to \emph{comm} and \emph{solver} and therefore do not appear in Figure~\ref{fig:micro:refill:bk}. As $L$ increases, the time spent in finding new treaties also increases. However, since the algorithm is able to find better treaties with larger values of $L$, the percentage of transactions which actually cause a treaty violation decreases as shown in Figure~\ref{fig:micro:refill:sync}. As expected, using a larger value for REFILL leads to a significantly lower synchronization ratio and higher throughput as shown in Figures~\ref{fig:micro:refill:tps} and \ref{fig:micro:refill:sync}.

\textbf{Varying the number of items accessed} For this experiment, we modify our microbenchmark transaction to order more than one item at a time. We vary the number of items purchased in a single transaction from 1 to 5. The latency CDF for the experiment is shown in Figure~\ref{fig:microbm:item:ltc}. As expected, as the transactions purchase more items, the probability of synchronization increases since it is more likely that the treaty governing any one of the purchased items is violated. This manifests itself in the downward shift of the inflection point in the CDF. This inflection point differentiates between locally executed transactions and those which cause a treaty violation. Under 2PC, the system behavior is unaffected by increasing the number of items ordered per transaction.

\subsection{TPC-C setting}\label{subsec:appendix_tpcc_exp}

Finally, we give detailed results for the distributed TPC-C experiment introduced in Section~\ref{sec:tpcc}. Figure~\ref{fig:tpcc:exp1_new:tps} shows how increasing the value of $H$ (the percentage of transactions ordering ``hot'' items) affects transaction throughput. As the value of $H$ increases, the treaties corresponding to hot items are violated more often, thus throughput falls. This increase in treaty violations is also apparent in Figure~\ref{fig:tpcc:exp1_neq:sync} where the synchronization ratio increases with increasing values of $H$. In summary, in a realistic workload and deployment scenario, the homeostasis protocol far outperforms 2PC and achieves near optimal throughput without sacrificing consistency.

\end{document}